\documentclass[letterpaper,twocolumn,10pt]{article}
\usepackage{usenix}
\usepackage{amsmath,amssymb,amsthm}
\usepackage{booktabs}
\usepackage{graphicx}
\usepackage{tikz}
\usetikzlibrary{arrows.meta,positioning}
\usepackage{xspace}
\usepackage{enumitem}
\usepackage{array}
\usepackage{algorithm}
\usepackage{algpseudocode}
\usepackage{array}
\usepackage{filecontents}

\newcommand{\sys}[1]{\ensuremath{\mathsf{#1}}}
\newcommand{\VeilFL}{\sys{VeilFL}\xspace}
\newcommand{\PINE}{\sys{PINE}\xspace}
\newcommand{\ACORN}{\sys{ACORN}\xspace}
\newcommand{\secpar}{\lambda}
\newcommand{\negl}{\mathsf{negl}}
\newcommand{\Zp}{\mathbb{Z}_p}
\newcommand{\Zps}{\mathbb{Z}_p^{*}}
\newcommand{\field}{\mathbb{F}}
\newcommand{\getsr}{\xleftarrow{\$}}
\newcommand{\GOne}{\mathbb{G}_1}
\newcommand{\GTwo}{\mathbb{G}_2}
\newcommand{\GT}{\mathbb{G}_T}
\newcommand{\pair}{\mathsf{e}}
\newcommand{\gone}{g_1}
\newcommand{\gtwo}{g_2}
\newcommand{\hone}{h_1}
\newcommand{\htwo}{h_2}
\newcommand{\gacc}{g}
\newcommand{\roundlinkbase}{\widehat g}
\newcommand{\dst}[1]{\text{``#1''}}
\newcommand{\Hash}{\mathsf{H}}
\newcommand{\HashG}{\mathsf{H}_{\GOne}}
\newcommand{\HashToScalar}{\mathsf{H}_{\Zp}}
\newcommand{\Encode}{\mathsf{Encode}}

\newcommand{\Client}[1]{\mathsf{C}_{#1}}
\newcommand{\TargetClient}{\mathsf{C}^{\star}}
\newcommand{\Agg}[1]{\mathsf{A}_{#1}}
\newcommand{\AggZero}{\mathsf{A}_0}
\newcommand{\AggOne}{\mathsf{A}_1}
\newcommand{\Issuer}{\mathsf{Iss}}
\newcommand{\Directory}{\mathcal{D}}
\newcommand{\TraceComm}{\mathcal{T}}
\newcommand{\IdComm}{\mathcal{I}}
\newcommand{\PublicLog}{\mathcal{L}}
\newcommand{\IssuerPK}{X}
\newcommand{\Abar}{\bar A}
\newcommand{\Bbar}{\bar B}
\newcommand{\TracePK}{Y}
\newcommand{\IdentPK}{K}
\newcommand{\TraceHandle}{U}
\newcommand{\DirHandle}{J}
\newcommand{\roundlink}{N}
\newcommand{\CtOne}{R}
\newcommand{\CtTwo}{S}
\newcommand{\clientsecret}{u}
\newcommand{\PoK}{\mathsf{PoK}}
\newcommand{\Renr}{\mathcal{R}_{\mathsf{enroll}}}
\newcommand{\Rsub}{\mathcal{R}_{\mathsf{sub}}}
\newcommand{\Req}{\mathcal{R}_{\mathsf{eq}}}
\newcommand{\PolicyRecord}{\mathsf{Policy}}

\newcommand{\policyID}{\mathsf{policyID}}
\newcommand{\issuerKeyID}{\mathsf{issuerKeyID}}
\newcommand{\pineID}{\mathsf{pineID}}
\newcommand{\roundinfo}{\mathsf{round}}
\newcommand{\context}{\mathsf{ctx}}
\newcommand{\update}{\Delta}
\newcommand{\qvec}{\mathbf{x}}
\newcommand{\pinemsg}{M}
\newcommand{\msgdigest}{d}
\newcommand{\publicpart}{\mathsf{Pub}}
\newcommand{\sid}{\mathsf{sid}}
\newcommand{\fid}{\mathsf{fid}}
\newcommand{\faultthreshold}{q}
\newcommand{\pinecomp}{\varepsilon_{\mathsf{c}}}
\newcommand{\pinesound}{\varepsilon_{\mathsf{s}}}
\newcommand{\pinezk}{\varepsilon_{\mathsf{zk}}}
\newcommand{\Adv}{\mathsf{Adv}}
\newtheorem{theorem}{Theorem}
\newtheorem{lemma}[theorem]{Lemma}

\theoremstyle{definition}
\newtheorem{definition}{Definition}

\setlist[itemize]{leftmargin=*,itemsep=2pt,topsep=3pt}
\setlist[enumerate]{leftmargin=*,itemsep=2pt,topsep=3pt}

\tikzset{
  box/.style={draw, rounded corners=1.5pt, align=center, inner sep=4pt,
              font=\footnotesize, minimum height=0.68cm},
  arrow/.style={-{Latex[length=2mm]}, line width=0.5pt},
  darrow/.style={-{Latex[length=2mm]}, dashed, line width=0.5pt},
  note/.style={font=\scriptsize, align=center}
}
\newcolumntype{L}[1]{>{\raggedright\arraybackslash}p{#1}}

\begin{document}

\date{}

\title{\Large \bf VeilFL: Two-Stage Accountability for Anonymous Federated Learning}

\author{
{\rm Varesh Mishra \quad Aysajan Abidin \quad Bart Preneel}\\
COSIC, KU Leuven, Belgium
}

\maketitle

\begin{abstract}
Anonymous credentials can hide which registered client participates in federated learning, but their unlinkability also makes repeated misconduct difficult to associate. We present \VeilFL, a two-aggregator protocol that adds staged accountability to \PINE (USENIX Security '24). Each submission binds two private \PINE messages, a round-specific pseudonym, and a tracing pair to the same hidden credential secret. This design enables detection of duplicate submissions within a round without introducing a persistent identifier for normal participation. \VeilFL implements accountability in two stages. After a confirmed duplicate or joint \PINE rejection, a threshold tracing committee reveals a stable handle that links faults associated with the same credential. The client's registered identity is recoverable only after the public log records the requisite number of linked faults and a separate threshold committee authorizes identification. Accepted and unopened submissions remain unlinkable under the stated assumptions. This work establishes submission binding and duplicate detection, anonymity, fault non-frameability, controlled two-stage opening, and preservation of \PINE's privacy and validity guarantees. In the prototype implementation, creating a submission requires a median of $2.11$\,ms, while verification requires $3.01$\,ms. Batch verification reduces the cost to $1.86$\,ms per submission for a batch of $1{,}000$. The public part of each submission is $608$ bytes, with $1{,}248$ bytes of additional \VeilFL protocol data across the two aggregator submissions. Each opening stage completes within $4.3$\,ms for the different committee configurations.
\end{abstract}

\section{Introduction}\label{sec:intro}
Federated learning trains a shared model from updates computed over data that remains with each client~\cite{mcmahan17}. Individual gradients and model updates can nevertheless disclose information about local examples and, in some settings, permit their reconstruction~\cite{dlg}. Secure aggregation addresses this exposure by revealing an aggregate rather than the individual client updates~\cite{bonawitz17}. In the distributed aggregation model of \sys{Prio}, each client instead secret-shares its input among non-colluding servers, which learn the aggregate but not the individual inputs under the stated corruption bound~\cite{prio}.

Protecting the updates also prevents the aggregators from inspecting them directly. A malicious client can alter the submitted update rather than merely changing its local training data, enabling model-poisoning and backdoor attacks~\cite{backdoorfl,fang}. Private input validation restores a limited check without revealing the update. In particular, \PINE allows two non-colluding verifiers to check that a secret-shared, high-dimensional integer vector satisfies an exact Euclidean-norm bound~\cite{pine}. It enforces the announced bound on the encoded contribution and thereby limits one way in which a client can submit an unusually large update.

However, private validation does not address who is allowed to participate. A deployment may wish to accept updates only from registered clients while hiding which registered client joined a particular round. Anonymous credentials can provide this form of admission. Their ordinary unlinkability, however, makes it difficult to implement the accountability mechanism. If several submissions are rejected, the service cannot tell whether they came from several clients that each made one mistake or from one client using fresh presentations of the same credential.

For instance, consider a consortium whose registered members train a shared model.
Participation in a particular round may itself be sensitive, while repeated submissions outside the agreed rules may eventually require action by the consortium's governance. A persistent identifier would make such behaviour easy to recognise, but would also make ordinary participation linkable. 
Conversely, opening the client's identity after a single event may be too strong when an out-of-bound update can be accidental. The consortium instead needs to associate repeated faults with one unknown member before deciding whether identification is justified.

Existing systems provide different accountability outcomes.
Anonymous-blacklisting systems such as \sys{Nymble}, \sys{BLAC}, and \sys{WhiteCloak} can exclude a misbehaving anonymous user without revealing their identity~\cite{nymble,blac,whitecloak}. 
Other systems make identity disclosure visible or auditable~\cite{papr,ats}.
\VeilFL targets a different point. It first links confirmed faults without identifying the client and permits identity resolution only after the public log records the required number of linked faults.

\VeilFL adds this staged accountability to an unmodified \PINE check.
Accepted submissions remain on the normal aggregation path and are never opened. When the public log confirms a fault, a tracing committee can first reveal a stable handle that lets the log associate repeated faults without consulting the issuer's directory. Only after the log contains \(\faultthreshold\) distinct linked faults under one handle can a separately keyed identification committee derive the handle used by the issuer's directory. Under the stated committee-corruption bounds, the first stage does not reveal the directory entry, and accepted and unopened submissions remain unlinkable even after another fault from the same client has been identified.

Using \PINE introduces one important composition problem since the client sends a different private message to each verifier. Two independent credential presentations would not by themselves establish that these messages belong to the same anonymous submission. \VeilFL instead covers the ordered hashes of both messages in one BBS presentation and binds the accountability values to the same hidden credential secret. Section~\ref{sec:rationale} explains why these bindings are needed, and Section~\ref{sec:submission} gives the complete construction.

The high-dimensional norm relation remains inside \PINE, so the added credential proof has fixed size with respect to the model dimension. \VeilFL also introduces no client-to-client pairwise key or masking state, unlike \ACORN-based systems such as \sys{WhiteCloak}. The two fault types expose different evidence. A duplicate can be checked from the public submissions, whereas a \PINE rejection depends on private verifier messages and cannot be reproduced by the log. \VeilFL therefore treats matching signatures from the two protocol-compliant aggregators as the evidence for a rejection. The log verifies those signatures and the public submission, but it does not rerun the private \PINE check.

\paragraph{Contributions.}
We make the following contributions:
\begin{itemize}
  \item \emph{Two-stage accountability for anonymous aggregation.}
  We introduce an accountability policy in which confirmed faults can first be linked without identifying the client, while identity resolution requires \(\faultthreshold\) distinct linked faults and a separate committee. Accepted submissions do not enter the opening process.

  \item \emph{Binding anonymous admission to two-verifier validation.}
  We construct one presentation that binds both private \PINE messages, the round-specific duplicate check, and the tracing information to the same hidden credential secret, while leaving \PINE's norm relation unchanged.

  \item \emph{Security analysis.}
  We prove submission binding and duplicate detection, anonymity and unlinkability of unopened submissions, fault non-frameability, controlled two-stage opening, and preservation of \PINE's aggregation privacy and validity under the stated assumptions.

  \item \emph{Implementation and evaluation.}
   We implement \VeilFL and evaluate its local cryptographic cost and encoded protocol size. A client submission takes a median of $2.11$\,ms and verification takes $3.01$\,ms. Batch verification reduces the cost to $1.86$\,ms per submission for a batch of $1{,}000$. The public part of a submission is $608$ bytes, and both opening stages complete within $4.3$\,ms for the committee configurations we evaluate.
\end{itemize}

\section{Design Overview}\label{sec:design}
\VeilFL combines anonymous admission, two-verifier private validation, and staged accountability. These components do not compose directly. The two verifier messages must belong to one anonymous submission,
duplicate detection must not make ordinary participation linkable across rounds, and fault linking must remain separate from identity disclosure. 
We first explain these problems and then give the high-level structure of the resulting protocol. Section~\ref{sec:protocol} gives the complete construction.

\subsection{Design Challenges}\label{sec:rationale}
A direct combination of anonymous credentials, \PINE, and threshold opening leaves three problems.

\smallskip\noindent\emph{Binding both verifier messages.}
A \PINE client sends a different private message to each verifier.
Two independent credential presentations would show that each message came from a registered client, but would not establish that both messages belong to the same submission. 
A persistent client signature could provide this link, but would also make ordinary submissions linkable.

\smallskip\noindent\emph{Detecting duplicates within a round.}
Unlinkable credential presentations provide no stable value with which to detect repeated submissions. A persistent pseudonym would solve this problem but would reveal participation across rounds. Omitting a pseudonym altogether would allow the same credential to contribute multiple submissions in one round. The protocol therefore needs a value that is stable within one round but changes between rounds.

\smallskip\noindent\emph{Separating linking from identification.}
Opening directly to the issuer's directory handle would identify the client after the first confirmed fault. We instead need a first opening that is sufficient only to associate faults from the same credential, followed by a separate identification step once the public fault history satisfies the announced policy.

These choices must also leave the underlying validation unchanged.
\PINE should continue to check the high-dimensional update, while the accountability layer handles only the information needed to bind the two verifier messages to the anonymous submission.

\subsection{\VeilFL Overview}\label{sec:design-overview}
\begin{figure*}[t]
  \centering
  \includegraphics[width=\textwidth]{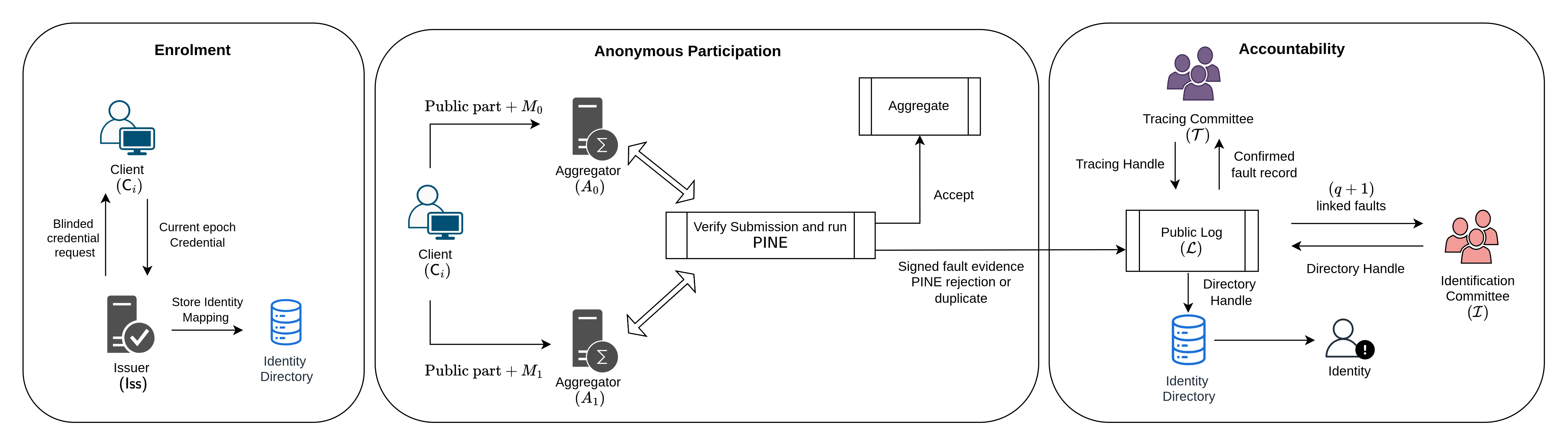}
  \caption{System architecture of \VeilFL. Clients obtain credentials and submit a common public part together with a private \PINE message to each aggregator. Accepted updates are aggregated, while confirmed faults enter a two-stage accountability path. Tier~1 reveals a tracing handle for linking faults, and Tier~2 derives a directory handle for identity resolution after the fault threshold is reached.}
  \label{fig:overview}
\end{figure*}

Figure~\ref{fig:overview} gives an overview of the three phases of \VeilFL - enrolment, anonymous participation, and accountability.
A client first obtains an anonymous credential from the issuer. During a training round, it prepares one private \PINE message for each aggregator and sends the same public part to both. The public part binds the two private messages to one credential, carries a round-specific value for duplicate detection, and contains the ciphertext used if the submission later becomes a confirmed fault. 
The aggregators first agree on the public part and exclude duplicate submissions, then execute the two \PINE verifier roles for the remaining submissions. Accepted shares are aggregated and do not enter the opening process.

Fault handling begins only after the log confirms either a duplicate or a joint \PINE rejection. 
The first opening gives the log a stable tracing handle under which faults from the same credential can be linked and counted. Once the log contains \(\faultthreshold\) distinct linked faults under that handle, a separate committee can perform the second opening needed for identity resolution. Thus, ordinary submissions remain unlinkable, while the fault path provides progressively more information only when the public record supports it. 

\section{Preliminaries}\label{sec:prelim}

\subsection{Groups, Encodings, and Notation}\label{sec:encoding}
Let \(p\) be prime, and let \((\GOne,\GTwo,\GT,\pair)\) be asymmetric bilinear groups of order \(p\). The BBS construction uses \(\gone,\hone,\htwo\in\GOne\) and \(\gtwo\in\GTwo\). The accountability construction uses \(\gacc\in\GOne\). All fixed \(\GOne\) bases are non-identity elements, and no discrete-log relation among them is known. We rely on decisional and computational Diffie-Hellman hardness in \(\GOne\).

The hash-to-group function \(\HashG\) maps directly to a non-identity element of \(\GOne\). In particular, the protocol does not reveal a scalar \(a\) such that \(\HashG(m)=\gacc^a\).

We write \(\Encode(m)\) for the fixed, length-delimited byte representation of a protocol object. It fixes the field order, lengths, and group and integer representations needed to interpret \(m\) unambiguously. \(\Hash\), \(\HashG\), and \(\HashToScalar\) are domain-separated hash functions to bytes, \(\GOne\), and \(\Zp\), respectively. All received group elements must use the required representation and pass group-membership checks before use. Table~\ref{tab:notation} lists the notation used most often.

\begin{table}[t]
\centering
\caption{Main notation.}
\label{tab:notation}
\footnotesize
\setlength{\tabcolsep}{4pt}
\renewcommand{\arraystretch}{1.08}
\begin{tabular}{@{}lp{0.68\columnwidth}@{}}
\toprule
Symbol & Meaning \\
\midrule
\(\policyID\) & policy identifier under which faults are counted \\
\(\issuerKeyID\) & identifier of the current BBS issuer key \\
\(\pineID\) & identifier of the \PINE parameters used in a round \\
\(\context\) & hash of the jointly signed round record \\
\(\pinemsg_b\) & private message sent to \(\Agg{b}\), containing its \PINE data and a private hash nonce \\
\(\msgdigest_b\) & hash of \(\pinemsg_b\), bound to verifier role \(b\) \\
\(\mathcal{B}\) & integer bound on \(\sum_i x_i^2=\|\qvec\|_2^2\) checked by \PINE \\
\(\roundlink\) & round pseudonym  \\
\(d\)      & \PINE vector dimension \\
\(\field\) & \PINE field \\
\((\CtOne,\CtTwo)\) & tracing pair \\
\(\TraceHandle\) & tracing handle  \\
\(\DirHandle\) & directory handle \\
\(\faultthreshold\) & linked faults required before identification \\
\bottomrule
\end{tabular}
\end{table}

\subsection{Short BBS Signatures}\label{sec:bbs-prelim}
We use the short BBS signature analysed by Tessaro and Zhu~\cite{tz23}. For readability, we write the two-message form. Let the issuer hold \(x\getsr\Zps\) and publish \(\IssuerPK=\gtwo^x\).
For hidden messages \((\clientsecret,s)\), define
\[
 M=\gone\hone^{\clientsecret}\htwo^s.
\]
A BBS credential is a pair \((A,e)\), where \(e\in\Zp\setminus\{-x\}\) and
\[
 A=M^{1/(x+e)}.
\]
The credential verifies when
\begin{equation}
 \pair\!\left(A,\IssuerPK\gtwo^e\right)
 =
 \pair\!\left(M,\gtwo\right).
 \label{eq:bbs-credential}
\end{equation}

\VeilFL uses blind issuance during enrolment, as described in Section~\ref{sec:enrol}. For presentation, the holder does not reveal \((A,e)\), since doing so would provide a stable value across presentations. Instead, it samples \(r_1,r_2\getsr\Zps\), sets \(r_3=r_2^{-1}\), and computes
\[
\begin{aligned}
 D      &= M^{r_2},\\
 \Abar  &= A^{r_1r_2},\\
 \Bbar  &= D^{r_1}\Abar^{-e}.
\end{aligned}
\]
Fresh \(r_1\) and \(r_2\) produce new presentation values each time.
For a valid credential, these values satisfy the public pairing check
\begin{equation}
 \pair(\Abar,\IssuerPK)
 =
 \pair(\Bbar,\gtwo).
 \label{eq:bbs-presentation-pairing}
\end{equation}
The holder also proves knowledge of
\((e,r_1,r_3,\clientsecret,s)\) satisfying
\[
 \Bbar = D^{r_1}\Abar^{-e} \quad \text{and} \quad 
 \gone = D^{r_3}
          \hone^{-\clientsecret}
          \htwo^{-s}.
\]
Together with Eq.~\eqref{eq:bbs-presentation-pairing}, these relations form the randomised BBS message-signature presentation. We rely on its proof-of-knowledge and zero-knowledge properties from Tessaro and Zhu~\cite{tz23}. Appendix~\ref{app:submission-pok} gives the concrete submission proof used by \VeilFL.

\subsection{PINE validation}\label{sec:pine-prelim}
\PINE checks a bound on a secret-shared integer vector \(\qvec=(x_1,\ldots,x_d)\in\mathbb Z^d\) using two verifiers~\cite{pine}. We write its predicate as
\begin{equation}
  \|\qvec\|_2^2
  =\sum_{i=1}^{d}x_i^2
  \le \mathcal{B},
\label{eq:pine-bound}
\end{equation}
where the comparison is over the integers. The dimension \(d\), bound \(\mathcal{B}\), \PINE field \(\field\), fixed-point encoding, coordinate range, and security parameters are fixed by the \PINE-parameter record of Section~\ref{sec:records}. 
The field \(\field\) is separate from the scalar field \(\Zp\) used by the credential and opening constructions.

The selected field and protocol parameters satisfy the conditions required by \PINE for the announced bound. Although verification is performed over \(\field\), \PINE's wraparound check ensures that Eq.~\eqref{eq:pine-bound} is a statement about the encoded integer vector rather than only its value modulo the field size.

We consider the non-interactive statistical \PINE construction. The client completes the prover computation locally and obtains one verifier-specific prover message for each verifier before transmitting the \VeilFL submission. If this local proof-generation step aborts, \VeilFL sends no submission. For an honest in-bound vector, conditioned on successful proof generation and complete delivery, the protocol-compliant verifiers obtain \PINE's common accepting output.

We write \(\pi_b^{\mathsf{PINE}}\) for the verifier-specific prover data produced for verifier \(b\). Together with the additive share \(\qvec_b\), this is the client data processed by that \PINE verifier.
The verifiers subsequently perform \PINE's prescribed simultaneous verifier-to-verifier exchange to obtain the common validation result.
We denote by \(\pinecomp\) the completeness error of the selected \PINE profile. For an honest in-bound input, the local proof-generation step aborts with probability at most \(\pinecomp\). For the final statistical \PINE construction, the end-to-end guarantee uses the same error parameter for completeness and for the statistical distributed-zero-knowledge distance, so we set
\(
  \pinezk=\pinecomp.
\)
For a vector satisfying the \PINE input condition, one verifier's view can therefore be simulated up to statistical distance \(\pinezk\).
We use \(\pinesound\) for the end-to-end soundness error assigned to the selected non-interactive \PINE profile, including the finite-field soundness term and any adjustment required by the distributed Fiat-Shamir transformation.

Section~\ref{sec:submission} describes how \VeilFL binds the two verifier messages to an anonymous submission without changing this \PINE predicate or verifier procedure.

\section{System and Security Model}\label{sec:model}
We use three terms throughout the text. A submission is \emph{complete} when both aggregators hold the same valid public part and each has received the private message matching its advertised hash value. A fault is \emph{confirmed} when the log accepts a rejection or duplicate record. 
It becomes a \emph{linked fault} only after Tier~1 opens its tracing pair and the log associates the record with a tracing handle. Only linked faults count towards the identification threshold.

\subsection{Parties}\label{sec:entities}
We consider the following parties in the \VeilFL protocol and describe
their respective roles below.
\begin{enumerate}[label=\textbf{P\arabic*.}, ref=P\arabic*, leftmargin=*]
  \item \emph{Clients} \((\Client{i})\) are registered participants. In each round, a client prepares one private \PINE message for each aggregator and sends the same public part to both.

  \item \emph{Aggregators} (\(\AggZero\) and \(\AggOne\)) publish joint round records, verify submissions, execute the two \PINE verifier roles, and combine only the shares of accepted submissions. Each holds a signing key for round and fault records.

  \item The \emph{issuer} \((\Issuer)\) authenticates eligible accounts, issues BBS credentials, and maintains the private directory \(\Directory\), which maps directory handles to registered identities.

  \item The \emph{tracing committee} \((\TraceComm)\) opens a confirmed fault to a stable tracing handle.

  \item The \emph{identification committee} \((\IdComm)\) derives the corresponding directory handle only for a request supported by \(\faultthreshold\) linked faults.

  \item The \emph{public log} \((\PublicLog)\) records joint round records, confirmed faults, opening shares, links to tracing handles, and identification requests. Parties verify its entries against signed checkpoints.
\end{enumerate}

\subsection{Public Records and Round Context}\label{sec:records}
We use three separate public records, so that rotating the issuer key or changing the \PINE parameters does not start a new fault history.

The \emph{policy record} fixes the task's accountability configuration. This includes the aggregator verification keys, the tracing and identification keys, committees, thresholds, share-verification information, the log and directory identifiers and the number \(\faultthreshold\) of linked faults required for identification. Its identifier is \(\policyID=\Hash(\dst{policy},\Encode(\PolicyRecord))\). Faults are counted under \((\policyID,\TraceHandle)\). Changing an opening key, committee, log configuration, or \(\faultthreshold\) therefore requires a new policy record and starts a new fault history.

An \emph{issuer-key record} declares the BBS public key valid under a policy and its validity interval. A \emph{PINE-parameter record} fixes the dimension, field, integer encoding and range, squared-norm bound, selected statistical-\PINE parameters and error bounds, and common reference string. The field and encoding must satisfy \PINE's conditions for interpreting the bound over the integers. The identifiers \(\issuerKeyID\) and \(\pineID\) are domain-separated hashes of their respective records. Rotating the issuer key or changing the \PINE parameters changes only the corresponding record identifier and preserves the existing fault history. A deployment instead creates a new policy record when such a change should reset that history.
For round \(t\), both aggregators sign
\begin{equation}
\begin{aligned}
\roundinfo_t=(&\policyID,\issuerKeyID,\pineID_t,t,\\
  &\mathsf{modelDigest}_t,\mathsf{window}_t).
\end{aligned}
\label{eq:round-record}
\end{equation}
The log accepts at most one jointly signed record for each policy and round number. The corresponding round context is
\begin{equation}
 \context_t=
 \Hash(\dst{round-context},\Encode(\roundinfo_t)).
\label{eq:context}
\end{equation}
Every submission proof and private-message hash is bound to this context.

\subsection{Threat Model and Channels}\label{sec:assumptions}
A probabilistic polynomial-time \((\mathsf{PPT})\) adversary controls any number of malicious clients and the corrupted committee members. We state our assumptions below.

\begin{enumerate}[label=\textbf{A\arabic*.}, ref=A\arabic*, leftmargin=*]
  \item \emph{Clients.}
  Any number of clients may be malicious, hold valid credentials, and   deviate arbitrarily. Honest clients follow the protocol and create at most one submission per round context.

  \item \emph{Aggregators.}
  The two aggregators are independent, protocol-compliant, honest-but-curious parties. They perform the public checks, \PINE, aggregation, and signing procedures as specified and do not collude.
  They exchange only the verifier-to-verifier information required by \PINE and may otherwise retain their local views. Update privacy and the interpretation of a joint \PINE rejection rely on this assumption.

  \item \emph{Issuer.}
  The issuer is honest-but-curious. It follows the enrolment, directory, renewal, and one-active-credential rules while retaining its transcripts. Eligibility is determined by an external registration process.

  \item \emph{Committees.}
  Fewer than the tracing threshold and fewer than the identification threshold are corrupted for the lifetime of a policy key. Corruption is static, and the two committee memberships are disjoint. Honest tracing members release shares only for confirmed faults, while honest identification members release shares only for a request supported by \(\faultthreshold\) distinct linked faults under one policy and tracing handle.

  \item \emph{Log.}
  The log is append-only, consistent, and available when a committee checks an opening request. It follows the record, uniqueness, and checkpoint rules specified by the protocol and provides one consistent view to all parties.
  
  \item \emph{Network and delivery.}
  The adversary may delay, replay, or drop messages. Client-to-aggregator channels are confidential, integrity-protected, and aggregator-authenticated; the aggregator-to-aggregator channel is confidential, authenticated, and integrity-protected. Network-level participation privacy additionally requires sender-anonymous client transport.
\end{enumerate}

\paragraph{Out of scope.}
Our threat model excludes malicious or colluding aggregators, a malicious issuer or log, and adversaries that corrupt either committee at or above its assumed corruption threshold. We further exclude Sybil registration, voluntary credential sharing, and harmful updates that remain within the announced PINE bound. The protocol also does not provide network-layer anonymity. IP addresses and other network metadata may be observable unless it is composed with a sender-anonymous communication mechanism, such as a mixnet.

\subsection{Security goals}\label{sec:goals}
In this section, we state the security goals that guide the design and analysis of \VeilFL. We later show how the protocol achieves each goal under the threat model above in Section~\ref{sec:security}.

\begin{enumerate}[label=\textbf{G\arabic*.}, ref=G\arabic*, leftmargin=*]

  \item\label{goal:anonymous} \textbf{Anonymous submissions.}
  An unopened honest submission does not reveal which eligible client produced it. This remains true after another submission from the same client has been traced or identified, subject to the threat model above and excluding identifying application or network metadata.

  \item\label{goal:binding} \textbf{Binding and duplicate detection.}
  Both private \PINE messages, the round pseudonym, and the tracing information are bound to one credential presentation. Repeated use of the same credential within one round is detected before aggregation.

  \item\label{goal:nonframe} \textbf{Fault non-frameability.}
  Except with negligible cryptographic probability, an honest client that submits at most once in a round cannot acquire a confirmed fault that it did not cause. An honest \PINE prover abort produces no submission and therefore cannot create a fault.

  \item\label{goal:opening} \textbf{Two-stage opening.}
  Before Tier 1, a sub-threshold coalition cannot recover the tracing handle from an unopened tracing ciphertext beyond information already revealed by authorised openings. After Tier~1, a sub-threshold identification coalition cannot determine the corresponding directory handle, and honest identification members act only on a request supported by \(\faultthreshold\) distinct linked faults.

  \item\label{goal:aggregation} \textbf{Aggregation privacy and validity.}
  Every client vector included in the aggregate satisfies the announced \PINE predicate except with probability \(\pinesound\) per submission. For an honest in-bound submission, the additional \VeilFL values preserve the selected \PINE single-verifier privacy guarantee up to \(\pinezk\) and negligible cryptographic error.

\end{enumerate}

\section{The VeilFL Protocol}\label{sec:protocol}

In this section, we describe the protocol steps in detail. It is first important to clarify the notation used in this section. 
We describe one client in one round and omit client and round indices unless two executions are compared. The verifier index \(b\in\{0,1\}\) and committee-member index \(j\) remain because they distinguish different roles. Thus \(\pinemsg_b\), \(\msgdigest_b\), \(\roundlink\), \((\CtOne,\CtTwo)\), \(\publicpart\), and \(\sid\) refer to one submission. In order to keep the notation readable, we use submission-specific notation rather than carrying client indices through every equation.

\subsection{Setup}\label{sec:setup}
Setup fixes the public group description
\[
 (p,\GOne,\GTwo,\GT,\pair)
\]
and the bases
\(
 \gone,\hone,\htwo,\gacc\in\GOne,
\) and \(\gtwo\in\GTwo\). It also fixes the domain-separated hash functions and byte representation used throughout the protocol. These values are global public parameters for the task and are not repeated in every round record. For each issuer-key record, the issuer samples \(x\getsr\Zps\) and publishes
\(
 \IssuerPK=\gtwo^x.
\)
The two aggregators independently generate EUF-CMA signature key pairs. The tracing and identification committees run independent verifiable distributed key-generation procedures~\cite{gjkr}. The tracing committee obtains shares \(y_j\) of a secret \(y\) and publishes
\[
 \TracePK=\gacc^y,
 \qquad
 \TracePK_j=\gacc^{y_j}.
\]
Independently, the identification committee obtains shares \(\kappa_j\) of a secret \(\kappa\) and publishes
\[
 \IdentPK=\gacc^\kappa,
 \qquad
 \IdentPK_j=\gacc^{\kappa_j}.
\]
The public keys \(\TracePK\) and \(\IdentPK\) must be non-identity elements. The issuer, tracing, and identification secrets are generated independently. All accountability public keys and share-verification values above are published in \(\GOne\). It is important to state that Setup publishes no \(\GTwo\) element whose exponent is \(y\), \(\kappa\), \(y_j\), or \(\kappa_j\).

The tracing key defines threshold ElGamal encryption over \(\GOne\) with base \(\gacc\)~\cite{cgs97}. It is an encryption of \(M\in\GOne\) with randomness \(\rho\in\Zp\). 

The public records of Section~\ref{sec:records} bind these values to the task and its accountability policy. Each issuer key is declared by an issuer-key record, and the selected \PINE parameters are named by a \PINE-parameter record. 
A round is processed only when its jointly signed round record refers to valid records under the current policy.

\subsection{Enrolment and Credential Renewal}\label{sec:enrol}
During first enrolment under \(\policyID\), the client authenticates to the issuer and samples \(\clientsecret\getsr\Zps\). It computes
\[
 \DirHandle=\IdentPK^{\clientsecret}.
\]
For the current issuer key \(\IssuerPK=\gtwo^x\), the client samples \(s\getsr\Zp\) and forms
\(
 C'=\hone^{\clientsecret}\htwo^s.
\)
It sends \((\DirHandle,C')\) and proves
\[
\Renr:\quad
\PoK\{(\clientsecret,s):
 C'=\hone^{\clientsecret}\htwo^s
 \ \wedge\ 
 \DirHandle=\IdentPK^{\clientsecret}\}.
\]
The proof statement includes \(\policyID\), \(\issuerKeyID\), the authenticated enrolment-session identifier, \(C'\), and \(\DirHandle\). Before accepting an enrolment or renewal, the issuer checks that \(\DirHandle\) is a valid non-identity element of \(\GOne\) and verifies the proof for \(\Renr\). At first enrolment, it additionally checks that \(\DirHandle\) is not assigned to another account and records
\[
  \Directory[\policyID,\DirHandle]=\mathrm{id}.
\]
Here \(\mathrm{id}\) denotes client's identity information. At renewal, it requires the same handle already stored for that account and enforces at most one active credential under each issuer key. The issuer samples
\(
 e\getsr\Zp\setminus\{-x\}
\)
and returns
\(
 A=(\gone C')^{1/(x+e)}
\)
together with \(e\). The client accepts only if \(A\neq 1_{\GOne}\) and Eq.~\eqref{eq:bbs-credential} holds.

Renewal uses the same \(\clientsecret\) and \(\DirHandle\), but a fresh \(s\) and the current issuer key. It therefore changes the credential without resetting the fault history under \(\policyID\). A new policy deliberately starts a new history and may allow the client to choose a fresh secret.

\subsection{Update Submission}\label{sec:submission}
A submission contains one private \PINE message for each aggregator and one common public part that binds both messages to the credential. 
Algorithm~\ref{alg:submit} summarises the client computation. We explain the values relevant to later verification here. We write the public statement authenticated by the submission proof as
\[
\begin{aligned}
 x_{\mathsf{sub}}=\bigl(&
 \IssuerPK,\TracePK,\context,
 \msgdigest_0,\msgdigest_1, \Abar,\Bbar,D,
 \roundlinkbase,\roundlink,
 \CtOne,\CtTwo
 \bigr).
\end{aligned}
\]
\begin{algorithm}[t]
\caption{Client submission in one round}
\label{alg:submit}
\begin{algorithmic}[1]
\Require update \(\update\); credential \((A,e)\) on
\((\clientsecret,s)\); signed record \(\roundinfo\); public parameters

\State verify both aggregator signatures and log inclusion for
\(\roundinfo\)

\State
\(\context \gets
 \Hash(\dst{round-context},\Encode(\roundinfo))\)

\State map \(\update\) to the integer vector \(\qvec\) used by \PINE

\State \(\qvec_0\getsr\field^d\);
\(\qvec_1\gets\qvec-\qvec_0\) in \(\field^d\)

\State run the selected non-interactive statistical \PINE prover
\If{the \PINE prover aborts}
  \State \Return without submitting
\EndIf
\State let
\(\pi_0^{\mathsf{PINE}}\) and \(\pi_1^{\mathsf{PINE}}\)
be the resulting verifier-specific prover data

\For{\(b \in \{0,1\}\)}
  \State
  \(\nu_b \getsr \{0,1\}^{\secpar}\);
  \(\pinemsg_b \gets
    (\nu_b,\qvec_b,\pi_b^{\mathsf{PINE}})\)
  \State
  \(\msgdigest_b \gets
    \Hash(\dst{pine-message},\context,b,
          \Encode(\pinemsg_b))\)
\EndFor

\State
\(\roundlinkbase \gets
 \HashG(\dst{round-pseudonym},\context)\)
\State
\(\roundlink \gets
 \roundlinkbase^{\clientsecret}\)

\State
\(\rho \getsr \Zps\);
\(\CtOne \gets \gacc^\rho\);
\(\CtTwo \gets
 \gacc^{\clientsecret}\TracePK^\rho\)

\State randomise the credential to obtain
\((\Abar,\Bbar,D)\)

\State generate \(\pi^{\mathsf{sub}}\) for \(\Rsub\) on public
statement \(x_{\mathsf{sub}}\)

\State
\(\publicpart \gets
(\context,\roundlink,\msgdigest_0,\msgdigest_1,
 \CtOne,\CtTwo,\Abar,\Bbar,D,\pi^{\mathsf{sub}})\)

\State
\(\sid \gets
 \Hash(\dst{submission},\mathsf{version},
       \Encode(x_{\mathsf{sub}}))\)

\State send \((\publicpart,\pinemsg_b)\) to
\(\Agg{b}\) for \(b\in\{0,1\}\)

\end{algorithmic}
\end{algorithm}

\paragraph{Private messages.}
The client maps its update \(\update\) to an integer vector \(\qvec\) using the fixed-point encoding and coordinate range fixed for the round. The validity guarantee concerns this encoded vector. \PINE checks
\[
  \sum_{i=1}^{d}x_i^2\le \mathcal{B}.
\]
The client represents \(\qvec\) in the \PINE field and samples additive shares
\[
\qvec_0\getsr\field^d \quad\text{and}\quad
\qvec_1\gets\qvec-\qvec_0 \in \field^d
\]
It then runs the selected non-interactive statistical \PINE prover locally. For an honest in-bound vector, proof generation may abort with probability at most \(\pinecomp\). This abort occurs before \(\publicpart\) or either \(\pinemsg_b\) is transmitted, so it produces no \VeilFL submission and no rejection record. If proof generation succeeds, the prover returns the verifier-specific data \(\pi_0^{\mathsf{PINE}}\) and \(\pi_1^{\mathsf{PINE}}\). For each verifier \(b\in\{0,1\}\), the client samples
\[
  \nu_b\getsr\{0,1\}^{\secpar}
\]
and forms
\(
  \pinemsg_b=(\nu_b,\qvec_b,\pi_b^{\mathsf{PINE}}).
\)
The pair \((\qvec_b,\pi_b^{\mathsf{PINE}})\) is the data processed by \PINE. The nonce \(\nu_b\) is added by \VeilFL and is used only when hashing the private message.

The client computes
\begin{equation}
 \msgdigest_b = \Hash(\dst{pine-message},\context,b,\Encode(\pinemsg_b)).
\label{eq:message-digest}
\end{equation}
The verifier index binds the message to its intended aggregator, while the fresh nonce prevents the public digest from serving as an offline test for guesses about the unseen message.

\paragraph{Public part.}
The client derives the round-specific base and pseudonym
\begin{equation}
 \roundlinkbase=\HashG(\dst{round-pseudonym},\context)
 \quad \text{and} \quad
 \roundlink=\roundlinkbase^{\clientsecret}.
\label{eq:round-pseudonym}
\end{equation}
The base changes with the round context, so the same credential gives the same pseudonym within one round without exposing a stable public pseudonym across rounds.

The tracing pair
\(
 (\CtOne,\CtTwo)
 =
 (\gacc^\rho,
  \gacc^{\clientsecret}\TracePK^\rho)
\)
encrypts the group element \(\gacc^{\clientsecret}\). Tier~1 therefore recovers the tracing handle directly, without computing a discrete logarithm.

The client randomises its credential as described in Section~\ref{sec:bbs-prelim}, obtaining \((\Abar,\Bbar,D)\). The verifier can check
\begin{equation}
 \pair(\Abar,\IssuerPK)=\pair(\Bbar,\gtwo).
\label{eq:presentation-pairing}
\end{equation}
The client then proves
\begin{equation}
\Rsub:\ \PoK\!\left\{
 (e,r_1,r_3,\clientsecret,s,\rho):\;
 \begin{aligned}
  &\Bbar=D^{r_1}\Abar^{-e}\\
  &{}\wedge\ \gone=D^{r_3}\hone^{-\clientsecret}\htwo^{-s}\\
  &{}\wedge\ \roundlink=\roundlinkbase^{\clientsecret}\\
  &{}\wedge\ \CtOne=\gacc^\rho\\
  &{}\wedge\ \CtTwo=\gacc^{\clientsecret}\TracePK^\rho
 \end{aligned}
\right\}.
\label{eq:rsub}
\end{equation}

The Fiat-Shamir challenge covers \(x_{\mathsf{sub}}\) together with the prover commitments. An accepting proof therefore binds the two private message digests, round pseudonym, and tracing ciphertext to the same credential presentation. Appendix~\ref{app:submission-pok} establishes the exact transcript and proves the shared-witness relation.

Finally, the client forms \(\publicpart\) and computes \(\sid\) from \(x_{\mathsf{sub}}\). The identifier includes the two PINE-message digests but excludes only the submission proof \(\pi^{\mathsf{sub}}\). 
Thus, regenerating a valid submission proof for the same statement yields the same identifier and is treated as a replay, whereas changing any value bound by the statement changes \(\sid\), except with negligible probability of a hash collision. The same \(\publicpart\) is sent to both aggregators, while each receives only its own private \PINE message.

\subsection{Verification and Aggregation}\label{sec:verification}
The verification has two phases. First, each aggregator checks its own received pair locally, then the two run a joint phase. Locally, \(\Agg{b}\) admits a pair only if all of the following hold.
\begin{enumerate}
  \item The signed round record and the referenced policy, issuer-key, and \PINE records are valid, and they produce the valid context.

  \item From that context, the aggregator recomputes
  \[
    \roundlinkbase
      =\HashG(\dst{round-pseudonym},\context).
  \]

  \item All received group elements have the required byte format and valid group membership. The presentation elements, pseudonym, and \(\CtOne\) satisfy the required non-identity checks.

  \item The pairing relation
  \(
    \pair(\Abar,\IssuerPK)=\pair(\Bbar,\gtwo)
  \)
  holds, and the proof for \(\Rsub\) verifies for the complete public statement.

  \item The received private message hashes to \(\msgdigest_b\) under Eq.~\eqref{eq:message-digest}.

  \item The aggregator reconstructs \(x_{\mathsf{sub}}\) from the received public-part fields, the referenced issuer and tracing keys, and the recomputed \(\roundlinkbase\), and checks
  \[
  \sid=
  \Hash(\dst{submission},\mathsf{version},
        \Encode(x_{\mathsf{sub}})).
  \]
  It also checks that this identifier has not already been processed. A retransmission, or another accepting proof for the same proof statement, has the same identifier and is treated as a replay.
\end{enumerate}

\paragraph{Batch verification.}
For multiple submissions using the same issuer key, the pairing check in item~4 can be batched using standard randomized batch verification~\cite{bellare1998batch}. For submissions indexed by \(i\), the verifier samples fresh \(\alpha_i\getsr\Zps\) and checks
\[
  \pair\!\left(
    \prod_i \Abar_i^{\alpha_i},
    \IssuerPK
  \right)
  =
  \pair\!\left(
    \prod_i \Bbar_i^{\alpha_i},
    \gtwo
  \right).
\]
All other submission checks, including verification of \(\pi^{\mathsf{sub}}\), are performed individually. We evaluate this optimization in Section~\ref{sec:evaluation}.

\paragraph{Completion and duplicate filtering.}
The two aggregators exchange \(\sid\) and the encoded public part and continue only if every field matches. The submission is then complete. After the submission window closes, they group complete submissions by \((\context,\roundlink)\). Any group containing more than one distinct submission identifier is excluded before \PINE is run and produces the single duplicate event defined in Section~\ref{sec:fault-records}.

\paragraph{PINE and aggregation.}
For every remaining submission, the aggregators execute the two \PINE verifier roles. The public-part and duplicate checks are complete before the \PINE execution begins. The aggregators then perform \PINE's verifier-to-verifier exchange as specified by \PINE. The two outgoing verifier messages are exchanged simultaneously, and neither message depends on the message sent by the other verifier. 
The submission identifier \(\sid\) is used only to associate this exchange with the correct concurrent submission. Under protocol-compliant execution, the two verifier roles obtain \PINE's common validation output. If that output is accept, each aggregator adds its vector share to the running total. If it is reject, they form the rejection record of Section~\ref{sec:fault-records}. 
If the verifier exchange fails to complete or no valid common output is obtained, the submission is excluded without creating a fault and the deployment may abort the round.
At finalisation, the aggregate shares reveal only
\[
 \sum_{i\in\mathsf{Acc}_t}\qvec_{i,t},
\]
where \(\mathsf{Acc}_t\) is the accepted set. No individual private message or vector share is published.

\subsection{Fault Records}\label{sec:fault-records}
This section describes how \VeilFL records publicly verifiable evidence for the two protocol-defined fault events. We now state again that only a complete submission can produce a fault record. We recognise two cases:

\begin{enumerate}[label=\textbf{F\arabic*.}, ref=F\arabic*, wide]
    \item\label{fault:pine-rejection} \emph{PINE rejection.}
    If \PINE's common validation output is reject for a complete, non-duplicate submission, aggregator \(\Agg{b}\) signs
    \[
     m_{\mathsf{rej}}=\Encode(\dst{fault-rejection},
                   \policyID,\context,\sid).
    \]
    The record contains the public part and both signatures on the same message. The log can verify the submission proof and the signatures, but it cannot rerun \PINE without the two private messages. The record therefore shows that the two designated aggregators attested to the same common private rejection. One signature, a withheld signature, or a verifier exchange that does not produce a valid common output is not a fault.
    
    \item\label{fault:duplicate} \emph{Duplicate submission.}
    Consider all complete submissions with the same \((\context,\roundlink)\). If the group contains at least two distinct submission identifiers, let \(\sid_1<\sid_2\) be the two smallest identifiers in byte order, with corresponding public parts \(\publicpart\) and \(\publicpart'\). 
    All submissions in the group are excluded, and the group produces at most one duplicate record. After each aggregator has verified the corresponding private message for both submissions, it signs
    \[
     m_{\mathsf{dup}}=\Encode\!(\dst{fault-duplicate},
     \!\policyID,\context,\roundlink,\sid_1,\sid_2).
    \]
    The record contains both public parts and both aggregator signatures. Anyone can recompute \(\sid_1\) and \(\sid_2\) from the two submission-proof statements, verify that \(\sid_1\neq\sid_2\), verify both credential presentations, and check that both statements contain the same \(\context\) and \(\roundlink\). The signatures additionally attest that both submissions were complete at both aggregators. A replay, or another accepting proof for the same statement, has the same submission identifier and does not create a duplicate record.
\end{enumerate}

The log assigns the identifiers
\begin{align*}
 \fid_{\mathsf{rej}}
   &=\Hash(\dst{fault-rejection},\policyID,\context,\sid),
   \\
 \fid_{\mathsf{dup}}
   &=\Hash(\dst{fault-duplicate},\policyID,\context,\roundlink).
\end{align*}
It stores at most one rejection record per \(\sid\) and at most one duplicate record per \((\context,\roundlink)\). A duplicate record names \(\sid_1\) as the selected submission whose tracing information will be opened. Valid proofs with the same round pseudonym imply that both public parts use the same hidden secret. A record becomes a confirmed fault after the log verifies these checks. Confirmation alone does not increase any tracing handle's count. The record contributes only after Tier~1 links it to a tracing handle. An honest \PINE prover abort produces no complete submission and therefore cannot create a fault record.

\subsection{Tier 1: Fault Linking}\label{sec:tier1}
For a confirmed fault record \(F\), the log forms a tracing request containing \(\policyID\), \(\fid\), the selected \(\sid\), the relevant pair \((\CtOne,\CtTwo)\), and a signed checkpoint. An honest tracing member from \(\TraceComm\) verifies the record, its inclusion, and the selected public part. The tracing committee member \(j\) then publishes
\[
 W_j=\CtOne^{y_j}
\quad\text{and}\quad
 \PoK\{(y_j):
  \TracePK_j=\gacc^{y_j}
  \ \wedge\ 
  W_j=\CtOne^{y_j}\}.
\]
The Fiat-Shamir challenge includes the policy, committee and member identifiers, fault and submission identifiers, checkpoint, and ciphertext. A valid proof therefore cannot be relabelled as a share for another request. For a valid tracing quorum \(T_{\mathsf{tr}}\), the log combines the shares and obtains
\begin{equation}
 \TraceHandle
 =\frac{\CtTwo}
        {\prod_{j\in T_{\mathsf{tr}}}
             W_j^{\lambda_{j,T_{\mathsf{tr}}}}}
 =\gacc^{\clientsecret}.
\label{eq:tier1-open}
\end{equation}
The log then associates \(F\) with \((\policyID,\TraceHandle)\). The record is now a linked fault and contributes once to that handle's count. The tracing handle remains stable across rounds and issuer-key rotation under the same policy. We state that it is not the key used by the issuer's directory.

\subsection{Tier 2: Identification}\label{sec:tier2}
Once the log contains at least \(\faultthreshold\) distinct linked faults under \((\policyID,\TraceHandle)\), it may create
\[
 \mathsf{ReqID}=(\policyID,\TraceHandle,
        \mathcal{F},\mathsf{checkpoint}),
 \qquad |\mathcal{F}|=\faultthreshold,
\]
where \(\mathcal{F}\) is the set of fault identifiers supporting the request. An honest identification member from \(\IdComm\) verifies that every record in \(\mathcal{F}\) is distinct, confirmed, opened with valid Tier~1 shares, linked to the same handle and policy, and included in the checkpoint. It then publishes
\[
 Z_j=\TraceHandle^{\kappa_j}
\quad\text{and}\quad
 \PoK\{(\kappa_j):
  \IdentPK_j=\gacc^{\kappa_j}
  \ \wedge\ 
  Z_j=\TraceHandle^{\kappa_j}\}.
\]
The challenge binds the complete request and committee identity. Combining a valid identification quorum gives
\begin{equation}
 \DirHandle
 =\prod_{j\in T_{\mathsf{id}}}
      Z_j^{\lambda_{j,T_{\mathsf{id}}}}
 =\TraceHandle^\kappa
 =\IdentPK^{\clientsecret}.
\label{eq:tier2-open}
\end{equation}
The issuer resolves \(\Directory[\policyID,\DirHandle]\) and returns the registered identity together with the request and its public evidence. We state again that accepted submissions never need to be opened, and a directory handle is not produced at Tier~1.

\section{Security Analysis}\label{sec:security}
We now show that \VeilFL satisfies the security goals in Section~\ref{sec:goals} under the threat model in Section~\ref{sec:assumptions}. Appendix~\ref{app:pok} establishes the properties of the protocol's proofs of knowledge, while Appendix~\ref{app:security-games} gives the formal security experiments and detailed arguments for anonymity, non-frameability, and controlled opening.

\subsection{Cryptographic assumptions}\label{sec:security-setting}
Our proofs treat each domain-separated use of \(\Hash\), \(\HashG\), and \(\HashToScalar\) as an independent random oracle. For message binding, only collision resistance of \(\Hash\) over the protocol's fixed encoding is required. Let \(q_H\) denote the adversary's total number of queries to these random oracles. 
For BBS, we rely on knowledge soundness of the selected randomised message-signature presentation~\cite{tz23}. Credential admission under committed-message issuance additionally relies on the one-more unforgeability result of Tessaro and Zhu in the algebraic group model~\cite{tz23}. We assume DDH and CDH hardness in \(\GOne\), EUF-CMA security of the aggregator signatures, and secure threshold keys with correct reconstruction and security against subthreshold coalitions. 
For the tracing key, this includes threshold-ElGamal IND-CPA security. We also use the completeness, soundness, common-output, and distributed zero-knowledge guarantees of the selected \PINE profile from Section~\ref{sec:pine-prelim}.

\subsection{Binding and duplicate detection}\label{sec:security-binding}

\begin{lemma}[Credential continuity]\label{lem:credential-continuity}
Fix a policy identifier. Every credential issued or renewed for one honest account contains the same client secret \(\clientsecret\) and gives the same directory handle \(\DirHandle=\IdentPK^{\clientsecret}\), except with negligible probability.
\end{lemma}

\begin{proof}
By Lemma~\ref{lem:enrol-pok}, knowledge soundness of
\(\pi^{\mathsf{enr}}\) gives values \((\clientsecret,s)\) such that \(C'=\hone^{\clientsecret}\htwo^s\) and \(\DirHandle=\IdentPK^{\clientsecret}\). The issuer stores this handle for the authenticated account. Renewal is accepted only for that stored handle and for a new commitment proved to contain the same exponent.
Since \(\IdentPK\) is a non-identity element of a prime-order group, \(\IdentPK^{\clientsecret}=\IdentPK^{\clientsecret'}\) implies \(\clientsecret=\clientsecret'\bmod p\). Blind issuance then signs the new commitment. Fresh blinding and a new issuer key change the credential, but not the secret or directory handle.
\end{proof}

\begin{theorem}[Submission binding and duplicate detection]\label{thm:binding}
Every complete submission that passes the public checks has a valid BBS credential witness \((\clientsecret,s)\) and some \(\rho\in\Zps\) such that
\(
 \roundlink=\roundlinkbase^{\clientsecret},\,
 (\CtOne,\CtTwo)
   =(\gacc^\rho,
     \gacc^{\clientsecret}\TracePK^\rho),
\)
and each \(\msgdigest_b\) matches the private message received by \(\Agg{b}\). Two distinct such statements with the same \((\context,\roundlink)\) use the same \(\clientsecret\) and, except with negligible probability of a hash collision, are detected as one duplicate group and excluded before \PINE. A replay of the same statement is ignored.
\end{theorem}

\begin{proof}[Proof (sketch)]
Knowledge soundness of the submission proof extracts one tuple \((e,r_1,r_3,\clientsecret,s,\rho)\). Together with the public pairing check, the standard BBS presentation theorem implies that \((\clientsecret,s)\) belongs to a valid BBS credential witness, while the remaining equations use the same \(\clientsecret\) in the pseudonym and tracing ciphertext.

Each aggregator also checks the hash of the private message it received. Replacing that message without changing the advertised digest therefore requires a collision in \(\Hash\). Finally, if two distinct complete submissions have the same context and pseudonym, then
\(
 \roundlinkbase^{\clientsecret}
 =
 \roundlinkbase^{\clientsecret'}.
\)
Because \(\roundlinkbase\neq1_{\GOne}\) and \(\GOne\) has prime order, \(\clientsecret=\clientsecret'\bmod p\). The submission identifier is a collision-resistant hash of the proof statement, excluding the proof bytes, so distinct statements have distinct identifiers except with negligible probability. The duplicate rule excludes the complete group before \PINE is run.
Replaying or re-proving the same statement has the same identifier and is ignored, and the log's uniqueness rule creates at most one duplicate record for the group.
\end{proof}

\subsection{Submission anonymity}\label{sec:security-anonymity}
We summarise the anonymity experiment and give the proof sketch here. Appendix~\ref{app:anonymity} gives the formal experiment and complete hybrid argument. 
The adversary receives the issuer's retained enrolment and signing state, one aggregator's entire local view, sub-threshold committee shares, and both clients' tracing and directory handles. The challenge uses the same admissible vector in a fresh context and cannot be opened.

\begin{theorem}[Submission anonymity and unlinkability]\label{thm:anonymity}
The adversary has only negligible advantage in identifying which of the two challenge clients produced the unopened submission.
\end{theorem}

\begin{proof}[Proof (sketch)]
The argument uses four standard hybrids. First, we sample the randomised BBS presentation directly and simulate the submission proof. For fresh nonzero BBS randomisers, \((\Abar,D)\) is uniform over two non-identity elements of \(\GOne\), while \(\Bbar=\Abar^x\). 
Therefore this has the honest presentation distribution. Next, we replace the digest of the unseen private message by a random value. The fresh private nonce in that message makes an earlier random-oracle query to the exact hash input occur with probability at most \(q_H2^{-\secpar}\).

We then replace the tracing ciphertext by an encryption of an independent group element, using threshold-ElGamal IND-CPA security against a sub-threshold tracing coalition. Finally, we replace the round pseudonym by a random group element. Even if the adversary already knows \(\gacc^{\clientsecret}\) or \(\IdentPK^{\clientsecret}\), testing whether either has the same exponent as the round pseudonym is a DDH test in \(\GOne\). The final view is independent of the challenge client.
\end{proof}

\subsection{Fault non-frameability}\label{sec:security-frame}
We summarise the non-frameability experiment and give the proof sketch here. Appendix~\ref{app:frame} gives the formal experiment, full reduction, and case analysis. The adversary wins if it causes the log to confirm a rejection or duplicate that opens to an honest target's tracing handle.
\begin{theorem}[Fault non-frameability]\label{thm:nonframe}
For any PPT adversary,
\[
  \Pr\!\left[
    \mathsf{Exp}^{\mathsf{frame}}_{\VeilFL,\mathcal A}(\secpar)=1
  \right]
  \le \negl(\secpar).
\]
\end{theorem}

\begin{proof}[Proof (sketch)]
Consider a confirmed fault whose Tier~1 opening equals \(\TraceHandle^{\star}\). If its selected public part is unauthorised, opening correctness and Theorem~\ref{thm:binding} reduce this case to discrete-log hardness. A reduction uses its discrete-log challenge as the target tracing handle, simulates honest target submissions without knowing \(\clientsecret^{\star}\), and extracts
\(
  \clientsecret^{\star}
  =\log_{\gacc}\TraceHandle^{\star}
\)
from an unauthorised accepting submission. A duplicate contains at least one unauthorised submission because the honest target submits at most once per context. By Theorem~\ref{thm:binding}, this submission uses the same client secret as the target submission and therefore reduces to the preceding case.
Finally, an authorised submission cannot produce a rejection fault. Conditioned on successful \PINE proof generation and complete delivery, the protocol-compliant verifiers accept. Altering a private message under its advertised digest requires a hash collision, while fabricating a confirmed record without the required aggregator signatures requires a signature forgery. Appendix~\ref{app:frame} gives the full reduction and case analysis.
\end{proof}

\subsection{Controlled opening}\label{sec:security-opening}
We give the proof sketch here. Appendix~\ref{app:opening-proof} gives the detailed argument.

\begin{theorem}[Controlled opening]\label{thm:opening}
A valid tracing quorum recovers \(\TraceHandle=\gacc^{\clientsecret}\), and a valid identification quorum recovers \(\DirHandle=\IdentPK^{\clientsecret}\). Before Tier~1, a sub-threshold tracing coalition cannot recover or distinguish the handle in an unopened ciphertext beyond information revealed by authorised openings. After Tier~1, a sub-threshold identification coalition cannot compute the corresponding directory handle or identify it among known candidates. Honest identification members release shares only after \(\faultthreshold\) distinct linked faults under the same policy and tracing handle.
\end{theorem}
By Lemma~\ref{lem:credential-continuity}, under a fixed policy the recovered directory handle is the handle stored by the issuer and is preserved across credential renewals.
\begin{proof}[Proof (sketch)]
For a valid tracing quorum \(T_{\mathsf{tr}}\),
\[
 \frac{\CtTwo}
 {\prod_{j\in T_{\mathsf{tr}}}W_j^{\lambda_j}}
 =
 \frac{\gacc^{\clientsecret}\TracePK^\rho}{\CtOne^y}
 =
 \gacc^{\clientsecret}.
\]
For a valid identification quorum \(T_{\mathsf{id}}\),
\[
 \prod_{j\in T_{\mathsf{id}}}Z_j^{\lambda_j}
 =
 \TraceHandle^\kappa
 =
 \IdentPK^{\clientsecret}.
\]

The Chaum-Pedersen proofs bind every opening share to the member's registered verification value. Before Tier~1, first apply the same presentation-simulation and round-pseudonym DDH hybrids as in Theorem~\ref{thm:anonymity}. After these hybrids, the tracing ciphertext is the remaining client-secret-dependent challenge value. Threshold ElGamal IND-CPA security then hides its plaintext from a sub-threshold coalition. After Tier~1, target pseudonyms can be simulated by programming each fresh round base as \(\gacc^{a_t}\) and computing \(\TraceHandle^{a_t}\), while the randomised BBS presentation is simulated independently. Computing a previously unknown \(\TraceHandle^\kappa\) from \((\gacc,\IdentPK,\TraceHandle)\) is therefore a CDH problem, while testing a candidate directory handle is a DDH problem.

Honest identification members release their shares only after verifying \(\faultthreshold\) linked faults and the exponentiation itself does not enforce this rule. A corrupted quorum is outside the threat model.
\end{proof}

\subsection{PINE composition}\label{sec:security-pine}

\begin{theorem}[Transfer of the PINE guarantees]\label{thm:pine-composition}
\VeilFL preserves the selected \PINE profile's guarantees:
\emph{(i)} every vector included in the aggregate was accepted by both \PINE verifier roles and violates Eq.~\eqref{eq:pine-bound} with probability at most \(\pinesound\) per submission
\emph{(ii)} an honest in-bound proof-generation attempt aborts with probability at most \(\pinecomp\), and conditioned on no abort and complete delivery, it is accepted and creates no fault record and
\emph{(iii)} one aggregator's view of an honest in-bound submission, including the unseen-message digest \(\msgdigest_{1-\beta}\), is simulatable with error at most \(\pinezk+q_H2^{-\secpar}+\negl(\secpar)\).
\end{theorem}

\begin{proof}[Proof (sketch)]
Theorem~\ref{thm:binding} ensures that the two messages processed by \PINE are exactly those whose digests appear in the accepted submission proof. As \VeilFL does not change the \PINE relation, parameters, or verifier algorithm, the selected profile's soundness guarantee applies directly.

For an honest in-bound vector, local \PINE proof generation aborts with probability at most \(\pinecomp\), before any submission is sent. Conditioned on successful generation and complete delivery, the \PINE verifiers accept and all honest \VeilFL procedure checks succeed.

For privacy, invoke the selected \PINE simulator on the exposed aggregator's public input, vector share, and permitted outputs, incurring error \(\pinezk\). Its own private-message digest is computed from the simulated message. The other digest contains an independent private nonce and can be replaced by a random-oracle value except with probability \(q_H2^{-\secpar}\). 
The credential presentation, pseudonym, and tracing ciphertext are simulated as in Theorem~\ref{thm:anonymity}. Hence the total simulation error is at most
\(
  \pinezk+q_H2^{-\secpar}+\negl(\secpar).
\)
\end{proof}
\section{Evaluation}\label{sec:evaluation}
We evaluate the cost of the main \VeilFL protocol operations, batched verification, controlled opening, and the additional communication introduced by the protocol.

\subsection{Experimental Setup}
We evaluate a Rust implementation of \VeilFL on an Intel Core Ultra 7 165H machine with 32\,GB of memory running Ubuntu 26.04 LTS. We use \texttt{blstrs}/\texttt{blst} for BLS12-381 operations and SHA-256 for hashing. The BBS-style credentials, zero-knowledge proofs, and threshold-opening procedures are implemented directly using these primitives. All benchmarks run locally and single-threaded using the Criterion benchmarking framework, with 200 samples for core and enrolment operations and 100 for opening and batch verification, except batch size 1,000, which uses 50. Core measurements use \(d=1{,}000\), producing private payloads of 16,000 and 2,048 bytes; their hashing cost is included, but the underlying \PINE computation is not. We report median latency and interquartile ranges. Measurements exclude setup, networking, FL training, model execution, storage, and the underlying \PINE cost.

\subsection{Core Protocol Cost}
Table~\ref{tab:eval-core} reports the cost of the main \VeilFL operations. Creating a client submission takes a median of $2.11$\,ms, while verifying a submission takes $3.01$\,ms. Client enrolment takes $2.07$\,ms and the issuer-side enrolment operation takes $0.77$\,ms.
\begin{table}[!h]
    \centering
    \caption{Latency of the main \VeilFL protocol operations. Values are median latency in milliseconds. P25 and P75 denote the 25th and 75th percentiles.}
    \label{tab:eval-core}
    \small
    \begin{tabular}{lccc}
        \toprule
        Operation & Median (ms) & $P_{25}$ (ms) & $P_{75}$ (ms) \\
        \midrule
        Client enrolment        & $2.07$ & $1.93$ & $2.25$ \\
        Issuer enrolment        & $0.77$ & $0.60$ & $0.87$ \\
        Client submission       & $2.11$ & $1.91$ & $2.61$ \\
        Submission verification & $3.01$ & $2.83$ & $3.17$ \\
        \bottomrule
    \end{tabular}
\end{table}
\paragraph{Communication cost.} The encoded public part of a \VeilFL submission is $608$ bytes. With the inclusion of the fresh $16$-byte nonce sent on each of the two client-to-aggregator links, \VeilFL adds
\(
    2(608 + 16) = 1{,}248 \text{ bytes}
\)
of protocol content across the two links. This value excludes the underlying \PINE messages and generic transport framing.

\subsection{Batch Verification}
\begin{figure}[t]
    \centering
    \includegraphics[width=0.90\columnwidth]{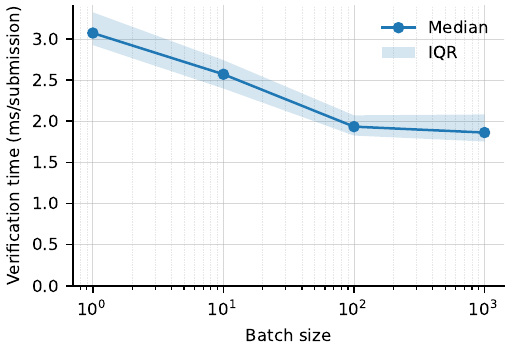}
    \caption{Amortized submission-verification cost as the batch
    size increases. The line shows the median and the shaded region
    shows the 25th-75th percentile range.}
    \label{fig:eval-batch}
\end{figure}

Figure~\ref{fig:eval-batch} shows the amortized verification cost when multiple submissions are processed using the batch-verification path. At batch size one, this path costs a median of $3.07$\,ms per submission, compared with $3.01$\,ms for ordinary single-submission verification in Table~\ref{tab:eval-core}. The amortized cost falls to $1.86$\,ms per submission at a batch size of $1{,}000$, a reduction of about 39\% relative to the batch-size-one measurement. Most of the reduction is already obtained at a batch size of 100, where verification costs $1.93$\,ms per submission.

\subsection{Controlled Opening}
\begin{figure}[t]
    \centering
    \includegraphics[width=.90\columnwidth]{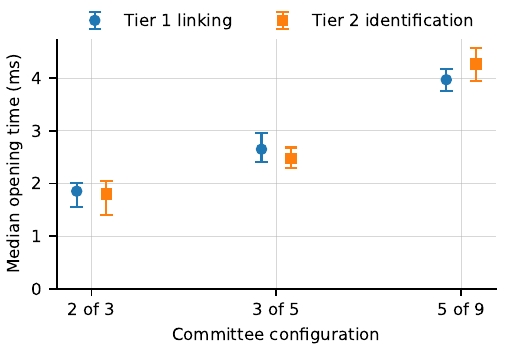}
    \caption{Latency of Tier~1 linking and Tier~2 identification
    for different threshold committee configurations. Points show
    median latency and error bars show the 25th-75th percentile
    range.}
    \label{fig:eval-opening}
\end{figure}

We next measure the cost of the two controlled-opening stages for different committee configurations. As shown in Figure~\ref{fig:eval-opening}, the cost increases with the number of shares required for opening. For a 2-of-3 committee, Tier~1 linking and Tier~2 identification take 1.85\,ms and 1.80\,ms, respectively. With a 5-of-9 committee, the corresponding costs are 3.97\,ms and 4.27\,ms. Thus, even the largest configuration we evaluate completes either opening stage in less than 4.3\,ms.

\section{Related Work}\label{sec:related}

Our work lies at the intersection of privacy-preserving FL aggregation and accountable anonymous participation. We review validated secure aggregation, anonymous authentication, and controlled linking and identification. Table~\ref{tab:compare} summarises selected related systems. These systems differ in their goals, validation methods, and trust assumptions.

\begin{table*}[!t]
\centering
\caption{Selected functional comparison with related systems.}
\label{tab:compare}
\scriptsize
\setlength{\tabcolsep}{3pt}
\renewcommand{\arraystretch}{1.12}

\begin{tabular}{@{}
L{0.155\textwidth}
L{0.125\textwidth}
L{0.125\textwidth}
L{0.155\textwidth}
L{0.155\textwidth}
L{0.185\textwidth}
@{}}
\toprule
\textbf{Property}
& \textbf{\PINE~\cite{pine}}
& \textbf{\ACORN~\cite{acorn}}
& \textbf{\sys{WhiteCloak}~\cite{whitecloak}}
& \textbf{\sys{PAPR}~\cite{papr}}
& \textbf{\VeilFL} \\
\midrule

\textbf{System model}
& 2 verifiers
& 1 server
& 1 server with \ACORN
& Credential system
& 2 aggregators \\

\textbf{Update validation}
& Exact encoded-integer \(\ell_2\) bound
& \(\ell_0,\ell_2,\ell_\infty\) predicates
& \ACORN predicates
& Not applicable
& Exact encoded-integer \(\ell_2\) bound via \PINE \\

\textbf{Anonymous participation}
& No
& No
& Yes
& Yes
& Yes \\

\textbf{Cross-round fault accountability}
& None
& None
& Anonymous blacklisting
& Not applicable
& Links confirmed faults across rounds \\

\textbf{Identity opening}
& Not provided
& Not provided
& Not provided
& After a logged request
& After \(\faultthreshold\) linked faults via a separate committee \\

\textbf{Client-to-client coordination}
& None
& Pairwise keys/masks
& Pairwise keys/masks
& Not applicable
& None \\

\bottomrule
\end{tabular}
\end{table*}

\paragraph{Private aggregation with input validation.}
Prio introduced distributed private aggregation in which clients secret-share their inputs among independent servers and provide proofs that the inputs satisfy validity conditions~\cite{prio}.
Bonawitz et al.\ give a practical single-server secure aggregation protocol using pairwise masks and dropout recovery~\cite{bonawitz17}. RoFL extends single-server secure aggregation with private checks on the \(\ell_2\) and \(\ell_\infty\) bounds of model updates~\cite{rofl}. EIFFeL supports general integrity checks and removes malformed updates from the aggregate while preserving update privacy~\cite{eiffel}. ELSA distributes trust across two servers, protecting individual updates as long as one server is honest while also supporting checks against malicious clients~\cite{elsa}. \ACORN supports several input constraints in a communication-efficient single-server design~\cite{acorn}. Its use of pairwise correlated keys requires checks on key consistency and mechanisms for recovering masking state when clients drop out.

\PINE uses two verifiers to check an exact Euclidean-norm bound on a secret-shared vector~\cite{pine}.
In this approach, the client sends a distinct private message to each verifier, and the verifiers exchange the information required to complete the check. \VeilFL retains this validation procedure and introduces anonymous admission and accountability for confirmed faults. A single credential proof authenticates the ordered digests of both private verifier messages and binds the round pseudonym and tracing ciphertext to the same hidden client secret. In combination with duplicate-submission rules, this mechanism enforces at most one accepted contribution per registered client per round and enables subsequent fault tracing to be linked to the same client.

\paragraph{Anonymous participation and accountability.}
Talwar et al.\ propose Samplable Anonymous Aggregation (SAA), which aggregates a randomly sampled subset of clients while hiding which clients contributed~\cite{saa}. Its Prio-based design supports anonymous, rate-limited participation and contribution checks.

\sys{WhiteCloak} studies accountability for anonymous clients in secure aggregation~\cite{whitecloak}.
It extends \ACORN with anonymous credentials and multilevel blacklisting, preventing excluded clients from returning under changed credentials while keeping their identities hidden. It uses Groth16 proofs for credential and blacklist checks. Because \ACORN uses pairwise correlated keys, \sys{WhiteCloak} also handles the seed material needed to attribute inconsistent keys and remove an excluded contribution.

\VeilFL supports a further response to confirmed faults. The faults can be linked and counted before a separate committee authorises identification. Here, the clients do not share pairwise masking keys, so rejecting one submission does not require removing masks shared with other clients. The submission proof connects the participation limit, both private verifier messages, and the encrypted fault handle, allowing the validation outcome to provide accountability at a later stage.

\paragraph{Anonymous authentication and identity disclosure.}
Group signatures allow an opening authority to identify the signer of a valid anonymous signature~\cite{bbs04,bsz}. Bounded-use anonymous authentication enables tracing when a user exceeds an allowed number of authentications~\cite{ktaa}. Anonymous-blacklisting systems such as Nymble and BLAC allow services to block misbehaving users while preserving their anonymity~\cite{nymble,blac}.

PAPR makes privacy revocation for anonymous credentials publicly auditable~\cite{papr}. It hides committee membership from the authority, requiring the authority to post a public request to contact the committee that can help disclose identity.
Kohlweiss and Miers' accountable metadata-hiding escrow allows users to check retrospectively whether they were traced~\cite{ats}. \VeilFL separates the round contribution limit from the decision to identify a client. The identification policy addresses distinct confirmed faults, including completed validation rejections. The fault rules prevent replays, repeated proofs of the same submission, and incomplete submissions from increasing the count used for identification.

\paragraph{Controlled linking of group signatures.}
Hwang et al.\ separate signature linking from signer identification using a BBS-family credential, encrypted handles, and a joint proof~\cite{hwang2013cl}. Slamanig et al.\ give a general method for adding controlled linking to suitable pairing-based group signatures~\cite{slamanig2014cl}. Their later work provides stable anonymous linking tokens and distributes their computation across several authorities for revocation checks~\cite{slamanig2017lbr}.

\VeilFL integrates these capabilities with round limits and confirmed private-validation faults. The first threshold opening reveals \(\TraceHandle\), which supports fault counting. The issuer's directory instead uses \(\DirHandle=\TraceHandle^\kappa\). Derivation of this directory handle requires the independent identification key \(\kappa\) and a second committee. Honest identification members release their shares only after checking at least \(\faultthreshold\) distinct linked faults. This separates fault-handle publication from directory lookup and ties identification to the recorded fault history.

\paragraph{Selective linking.}
Garms and Lehmann combine a BBS+ credential, an encrypted value derived from the credential secret, and a proof that both use the same secret~\cite{garms2019selective}. Their conversion procedure links selected batches of signatures while hiding the conversion inputs from the linking authority. It limits links across separate conversion requests, apart from information revealed by the signed messages. Their construction does not provide an identity-opening algorithm.

\VeilFL preserves a consistent anonymous handle for faults disclosed in different rounds under one policy.
A newly opened fault can therefore be added directly to the existing record without reconverting earlier signatures. A separate threshold stage then supports identity resolution from that record.
These operations support a continuing fault history and a separately authorised response to repeated faults.

\paragraph{BBS signatures and client proofs.}
In \emph{Revisiting BBS Signatures}, Tessaro and Zhu prove the security of short BBS signatures and give compact proofs of knowledge of a message-signature pair~\cite{tz23}. Related CFRG drafts describe BBS signatures, blind issuance, and context-specific pseudonyms~\cite{bbs-draft, blind-bbs-draft,bbs-pvl-draft}. 

\VeilFL extends Tessaro and Zhu's presentation techniques with equations for the round pseudonym and tracing ciphertext. It uses Schnorr-style proofs made non-interactive through the Fiat-Shamir transform.
Clients prove these group equations directly, without a general-purpose SNARK circuit for admission and tracing. The high-dimensional norm check remains in \PINE.

\section{Conclusion}\label{sec:conclusion}
\VeilFL lets registered clients take part in federated learning with two \PINE verifiers without making normal participation publicly linkable. The construction binds the two verifier messages, the credential, the round pseudonym, and the tracing ciphertext to the same hidden client secret while leaving \PINE's high-dimensional norm check unchanged. Fault handling uses two separate outputs. A confirmed rejection or duplicate can first be opened to \(\TraceHandle=\gacc^{\clientsecret}\), which lets the log count faults without consulting the issuer's directory. Only after \(\faultthreshold\) linked faults does the second committee derive \(\DirHandle=\IdentPK^{\clientsecret}\). Accepted and unopened submissions do not enter this process and remain outside the fault history.

\appendix
\section*{Ethical Considerations}
This work studies the design and evaluation of a cryptographic protocol for privacy-preserving and accountable federated learning. The research does not involve human subjects, user studies, personal data, private datasets, or interaction with operational federated learning systems. The evaluation is conducted using a research prototype and benchmark workload, without collecting information from real users or production environments. \VeilFL aims to improve participant privacy by separating a client's identity from its routine participation in federated learning, while maintaining a controlled accountability mechanism when a protocol-defined fault is established. This capability also creates deployment responsibilities. The tracing authorities must be subject to appropriate governance, and the protocol's threshold and trust assumptions must be upheld to prevent misuse of the accountability mechanism. The cryptographic guarantees do not address all potential sources of information leakage, such as network identifiers, timing and participation patterns, or information disclosed by model updates outside the secure aggregation protocol. These risks require separate operational and system-level mitigations in practical deployments.

\bibliographystyle{plainurl}
\bibliography{ref}

\section{Proofs of Knowledge}\label{app:pok}

To keep the equations readable, this appendix describes one execution and omits client, round, issuer-key, and committee indices. All exponents are in \(\Zp\). The Fiat-Shamir challenge includes the complete public statement in the protocol's fixed byte format. The hash-to-scalar function \(\HashToScalar\) is modelled as a random oracle.

\subsection{Enrolment proof}\label{app:enrol-proof}

The public statement is
\[
 x_{\mathsf{enr}}=(\policyID,\issuerKeyID,
   \mathsf{session},C',\DirHandle),
\]
and the relation is
\begin{equation}
 C'=\hone^{\clientsecret}\htwo^s,
 \qquad
 \DirHandle=\IdentPK^{\clientsecret}.
\label{eq:app-enrol-relation}
\end{equation}
The prover knows \((\clientsecret,s)\). It samples \(a_u,a_s\getsr\Zp\) and computes
\[
 T_1=\hone^{a_u}\htwo^{a_s},
 \qquad
 T_2=\IdentPK^{a_u}.
\]
It sets
\[
 c=\HashToScalar(\dst{enrol-proof},x_{\mathsf{enr}},T_1,T_2)
\]
and returns
\[
 \pi^{\mathsf{enr}}=(c,z_u,z_s),
 \quad
 z_u=a_u+c\clientsecret,
 \quad
 z_s=a_s+cs.
\]
The verifier reconstructs
\[
 \widehat T_1=\hone^{z_u}\htwo^{z_s}(C')^{-c},
 \qquad
 \widehat T_2=\IdentPK^{z_u}\DirHandle^{-c},
\]
and accepts if
\[
 c=\HashToScalar(\dst{enrol-proof},x_{\mathsf{enr}},
                 \widehat T_1,\widehat T_2).
\]

\begin{lemma}[Enrolment proof]\label{lem:enrol-pok}
In the random-oracle model, \(\pi^{\mathsf{enr}}\) has perfect completeness, knowledge soundness, and computational zero knowledge for \(\Renr\).
\end{lemma}

\begin{proof}
\emph{Completeness.}
Substituting the honest responses and Eq.~\eqref{eq:app-enrol-relation} gives
\[
 \widehat T_1
 =\hone^{a_u+c\clientsecret}\htwo^{a_s+cs}
  (\hone^{\clientsecret}\htwo^s)^{-c}
 =T_1
\]
and \(\widehat T_2=T_2\). The verifier therefore recomputes the prover's challenge.

\emph{Knowledge soundness.}
Suppose two accepting transcripts have the same commitments and distinct challenges \(c\neq c'\). The extractor computes
\[
 \clientsecret^*=\frac{z_u-z'_u}{c-c'},
 \qquad
 s^*=\frac{z_s-z'_s}{c-c'}.
\]
Taking the quotient of the two verification equations yields
\[
 C'=\hone^{\clientsecret^*}\htwo^{s^*},
 \qquad
 \DirHandle=\IdentPK^{\clientsecret^*}.
\]
Thus the extractor obtains a witness for \(\Renr\). The standard forking argument produces the two accepting transcripts from a successful Fiat-Shamir prover, except with negligible probability.

\emph{Zero knowledge.}
The simulator samples \(c,z_u,z_s\getsr\Zp\), reconstructs \(\widehat T_1,\widehat T_2\) with the verifier equations, and programs the random oracle at the resulting query to return \(c\). For a fixed challenge, the honest responses are uniform and thus, the simulated and honest transcripts have the same distribution. The programming-failure event is negligible in the random-oracle model.
\end{proof}

\subsection{Submission proof}\label{app:submission-pok}

Let \((A,e)\) be a BBS credential on \((\clientsecret,s)\), and write
\[
 M=\gone\hone^{\clientsecret}\htwo^s.
\]
Credential validity means
\[
 \pair(A,\IssuerPK\gtwo^e)=\pair(M,\gtwo).
\]
The selected three-point presentation samples \(r_1,r_2\getsr\Zps\), sets \(r_3=r_2^{-1}\), and computes
\[
 D=M^{r_2},
 \qquad
 \Abar=A^{r_1r_2},
 \qquad
 \Bbar=D^{r_1}\Abar^{-e}.
\]
The verifier checks
\begin{equation}
 \pair(\Abar,\IssuerPK)=\pair(\Bbar,\gtwo)
\label{eq:app-pairing}
\end{equation}
and the non-identity and subgroup conditions required by the selected BBS presentation.

For fixed public values \((\Abar,\Bbar,D,\roundlinkbase, \roundlink,\CtOne,\CtTwo)\), the representation relation is
\begin{align}
 \Bbar&=D^{r_1}\Abar^{-e},
 &\gone&=D^{r_3}\hone^{-\clientsecret}\htwo^{-s},
 \label{eq:app-sub-rel-1}\\
 \roundlink&=\roundlinkbase^{\clientsecret},
 &\CtOne&=\gacc^\rho,\quad
 \CtTwo=\gacc^{\clientsecret}\TracePK^\rho.
 \label{eq:app-sub-rel-2}
\end{align}
The public statement is \(x_{\mathsf{sub}}\) as defined in Section~\ref{sec:submission}.
The witness is
\[
  (e,r_1,r_3,\clientsecret,s,\rho).
\]
The prover samples
\(
  a_e,a_1,a_3,a_u,a_s,a_\rho\getsr\Zp
\)

and computes
\begin{align*}
 T_B&=D^{a_1}\Abar^{-a_e},
 &T_M&=D^{a_3}\hone^{-a_u}\htwo^{-a_s},\\
 T_N&=\roundlinkbase^{a_u},
 &T_R&=\gacc^{a_\rho},\\
 T_S&=\gacc^{a_u}\TracePK^{a_\rho}.
\end{align*}
It sets
\[
 c=\HashToScalar(\dst{submission-proof},x_{\mathsf{sub}},
                  T_B,T_M,T_N,T_R,T_S)
\]
and returns
\[
 \pi^{\mathsf{sub}}
 =(c,z_e,z_1,z_3,z_u,z_s,z_\rho),
\]
where
\begin{align*}
 z_e&=a_e+ce, &z_1&=a_1+cr_1,\\
 z_3&=a_3+cr_3, &z_u&=a_u+c\clientsecret,\\
 z_s&=a_s+cs, &z_\rho&=a_\rho+c\rho.
\end{align*}
The verifier reconstructs
\begin{align*}
 \widehat T_B&=D^{z_1}\Abar^{-z_e}\Bbar^{-c},\quad
 \widehat T_M=D^{z_3}\hone^{-z_u}\htwo^{-z_s}\gone^{-c},\\
 \widehat T_N&=\roundlinkbase^{z_u}\roundlink^{-c},\quad
 \widehat T_R=\gacc^{z_\rho}\CtOne^{-c},\quad
 \widehat T_S=\gacc^{z_u}\TracePK^{z_\rho}\CtTwo^{-c}.
\end{align*}
The verifier then computes
\[
 c'=\HashToScalar(
      \dst{submission-proof},x_{\mathsf{sub}},
      \widehat T_B,\widehat T_M,\widehat T_N,
      \widehat T_R,\widehat T_S)
\]
and accepts if and only if \(c'=c\).

\begin{lemma}[Submission proof]\label{lem:submission-pok}
For fixed randomised BBS presentation values that pass Eq.~\eqref{eq:app-pairing} and the required group checks, \(\pi^{\mathsf{sub}}\) has perfect completeness, knowledge soundness, and computational zero knowledge for the relation in Eqs.~\eqref{eq:app-sub-rel-1}-\eqref{eq:app-sub-rel-2}. Extraction yields a single \(\clientsecret^\star\) shared by the credential, round-pseudonym, and tracing-ciphertext relations.
\end{lemma}

\begin{proof}
\emph{Completeness.}
Substitution of the honest responses and Eqs.~\eqref{eq:app-sub-rel-1}-\eqref{eq:app-sub-rel-2} gives
\(\widehat T_B=T_B\), \(\widehat T_M=T_M\),
\(\widehat T_N=T_N\), \(\widehat T_R=T_R\), and
\(\widehat T_S=T_S\).

\emph{Knowledge soundness.}
Suppose two accepting transcripts have the same five commitments and distinct challenges \(c\neq c'\). Let \(\delta=(c-c')^{-1}\). The extractor sets
\begin{align*}
 e^*&=\delta(z_e-z'_e),
 &r_1^*&=\delta(z_1-z'_1),\\
 r_3^*&=\delta(z_3-z'_3),
 &\clientsecret^*&=\delta(z_u-z'_u),\\
 s^*&=\delta(z_s-z'_s),
 &\rho^*&=\delta(z_\rho-z'_\rho).
\end{align*}
Taking the quotient of the corresponding verification equations yields Eqs.~\eqref{eq:app-sub-rel-1}-\eqref{eq:app-sub-rel-2} for these extracted values. In particular,
\[
 \roundlink=\roundlinkbase^{\clientsecret^*}
 \quad\text{and}\quad
 \CtTwo=\gacc^{\clientsecret^*}\TracePK^{\rho^*}.
\]
The first two extracted equations, together with Eq.~\eqref{eq:bbs-presentation-pairing}, form the randomised BBS message-signature presentation analysed by Tessaro and Zhu~\cite{tz23}. By the knowledge-soundness property of that presentation, the extracted message values \((\clientsecret^*,s^*)\) correspond to a valid BBS credential except with negligible probability. The remaining equations already show that the same extracted value \(\clientsecret^*\) determines the round pseudonym and the plaintext encrypted by the tracing ciphertext.

\emph{Zero knowledge.}
The simulator samples \(c\) and all six responses uniformly, reconstructs the five commitments with the verifier equations, and programs the random oracle at the resulting query. For a fixed challenge, every honest response is uniform. Conditioned on the programmed point not having been queried earlier, the simulated proof has the same distribution as an honest proof for the fixed public statement. The programming-failure event is negligible in the random-oracle model.
\end{proof}

\subsection{Opening-share proofs}\label{app:share-pok}
Both committees use the same equality-of-discrete-logarithms proof. For public values \((G,Y,B,V)\), the relation is
\[
 \Req:\quad \PoK\{(\alpha):Y=G^\alpha\ \wedge\ V=B^\alpha\}.
\]
The prover samples \(a\getsr\Zp\), computes \(T_1=G^a\) and \(T_2=B^a\), and sets
\[
 c=\HashToScalar(\dst{opening-share},x_{\mathsf{open}},T_1,T_2),
 \qquad
 z=a+c\alpha.
\]
The verifier reconstructs
\[
 \widehat T_1=G^zY^{-c},
 \qquad
 \widehat T_2=B^zV^{-c}
\]
and checks the challenge.
For tracing, use \((G,Y,B,V,\alpha)=(\gacc,\TracePK_j,\CtOne,W_j,y_j)\) while for identification, use \((\gacc,\IdentPK_j,\TraceHandle,Z_j,\kappa_j)\). The statement \(x_{\mathsf{open}}\) includes the policy, committee, member, request, checkpoint, and relevant ciphertext or tracing handle.

\begin{lemma}[Opening-share proof]\label{lem:share-pok}
In the random-oracle model, the opening-share proof has perfect completeness, knowledge soundness, and computational zero knowledge for \(\Req\).
\end{lemma}

\begin{proof}
Completeness follows by direct substitution. 
From two accepting transcripts with the same commitments and challenges \(c\neq c'\), the extractor obtains \(\alpha=(z-z')/(c-c')\), and the quotient of the verification equations gives \(Y=G^\alpha\) and \(V=B^\alpha\). For zero knowledge, the simulator samples \(c,z\), reconstructs the commitments with the verifier equations, and programs the random oracle. Conditioned on the programmed point not having been queried earlier, this has the honest transcript distribution. The programming-failure event is negligible. This is the standard Chaum-Pedersen argument~\cite{chaum-pedersen}.
\end{proof}

\section{Security Experiments and Detailed Proofs}\label{app:security-games}
All experiments use the protocol algorithms and the threat model of Section~\ref{sec:assumptions}. Any additional leakage and challenge restrictions are stated in the corresponding experiment.

The challenger records the honest clients and the public parts returned by \(\mathsf{HonestSubmit}\). Only those public parts are \emph{authorised}. The oracle uses fresh credential-presentation, \PINE, and private-message randomness. If local \PINE proof generation aborts, it returns \(\bot\) and records no submission. It creates at most one authorised submission per client and round context.

A one-sided delivery is incomplete, while an altered private message fails its digest check except on a collision in \(\Hash\) and neither creates a fault. An honest retransmission reuses the same submission, and another accepting proof for the same statement has the same \(\sid\) and both are treated as replays.

For a non-replay adversarial submission, the proof statement differs from those of simulated honest submissions except on a collision in \(\Hash\). Since the statement is included in the Fiat-Shamir query, the standard forking extractor can keep the random-oracle points programmed for the simulated honest proofs fixed. Let \(q_H\) denote the adversary's total number of random-oracle queries. If an experiment-specific restriction is violated, that experiment returns \(0\).

\subsection{Submission anonymity}\label{app:anonymity}
This subsection defines the submission anonymity experiment and proves Theorem~\ref{thm:anonymity} through a sequence of hybrids. 
\begin{definition}[Submission anonymity experiment]\label{def:anonymity-game}
The experiment \(\mathsf{Exp}^{\mathsf{anon}}_{\VeilFL,\mathcal A}(\secpar)\) is defined as follows.
\begin{enumerate}
  \item The challenger runs setup, including both committee key-generation procedures, and gives \(\mathcal A\) all public records, the secret state of malicious clients, the entire local protocol view of one chosen aggregator, and a static subthreshold set of shares from each committee.
  
  \item \(\mathcal A\) names two honest accounts with the same public eligibility information. The challenger enrols them as \(\Client{0}\) and \(\Client{1}\), with secrets \(\clientsecret_0,\clientsecret_1\), and gives \(\mathcal A\) both issuer enrolment transcripts together with the issuer signing-key state used by those records and by the challenge round. To model prior opening in other rounds, it also gives
  \[
   \TraceHandle_i=\gacc^{\clientsecret_i},
   \qquad
   \DirHandle_i=\IdentPK^{\clientsecret_i},
   \qquad i\in\{0,1\}.
  \]
  
  \item Before and after the challenge, \(\mathcal A\) may request honest submissions from either client in any non-challenge context and receives the public part, the selected aggregator's private view, and the resulting public records. It may also run arbitrary malicious clients. It may not request a submission from either challenge client in the challenge context.
  
  \item \(\mathcal A\) chooses a fresh jointly signed round context and one admissible encoded vector \(\qvec^*\). The same vector and public attributes are used for either challenge client. The challenger samples \(b\getsr\{0,1\}\) and samples the challenge from the distribution of an honest complete submission by \(\Client{b}\), conditioned on successful \PINE proof generation. It then gives \(\mathcal A\) the common public part, the chosen aggregator's private \PINE message and verifier view, the common result, and all public log entries that the protocol would reveal.
  
  \item The challenge submission may not be sent to either opening procedure. Finally, \(\mathcal A\) outputs \(b'\). The experiment returns \(1\) if \(b'=b\).
\end{enumerate}
The advantage is
\[
 \Adv^{\mathsf{anon}}_{\VeilFL,\mathcal A}
 =\left|\Pr[b'=b]-\frac12\right|.
\]
\end{definition}

We condition the challenge on successful \PINE proof generation only to define the experiment; the client algorithm is unchanged and sends nothing if proof generation aborts. This conditioning does not favour
either client because both use the same vector, context, and \PINE parameters, and \PINE does not depend on the credential secret.
Both clients use the same vector so that the update itself does not reveal their identity. The context is fresh, so neither client has an earlier round pseudonym that can be compared with the challenge pseudonym.

\begin{proof}[Proof of Theorem~\ref{thm:anonymity}]
We use a sequence of experiments. Let \(\beta\in\{0,1\}\) denote the aggregator whose view is given to the adversary.

\textsf{Game 0.} \emph{(Real game).}
This is the real game as defined in  \(\mathsf{Exp}^{\mathsf{anon}}_{\VeilFL,\mathcal A}(\secpar)\).

\textsf{Game 1}: \emph{(Simulate the credential presentation).}
Replace the randomised BBS presentation and the submission proof by their simulators, not only in the challenge but in every honest submission returned for \(\Client{0}\) or \(\Client{1}\). For each such submission the challenger samples \(\Abar,D\getsr\GOne\setminus\{1_{\GOne}\}\), sets \(\Bbar=\Abar^x\), and simulates \(\pi^{\mathsf{sub}}\) using Lemma~\ref{lem:submission-pok}.

This replacement has the honest presentation distribution, except for negligible random-oracle programming failure. Since \(r_1,r_2\getsr\Zps\) are independent and nonzero, \((r_1r_2,r_2)\) is uniform over \((\Zps)^2\). For an accepted presentation, \(A\) and \(M\) are non-identity, so
\[
  (\Abar,D)=(A^{r_1r_2},M^{r_2})
\]
is jointly uniform over \((\GOne\setminus\{1_{\GOne}\})^2\), while \(\Bbar=\Abar^x\).

It is important to state that whenever a later hybrid changes a field of \(x_{\mathsf{sub}}\), the challenger reruns the proof simulator for the modified statement and recomputes \(\sid\) and any transcript values derived from it.

\textsf{Game 2}: \emph{(Hide the unseen private-message hash value).}
Replace \(\msgdigest_{1-\beta}\) by a uniform random-oracle value. Its honest preimage contains the independent nonce \(\nu_{1-\beta}\), so the adversary queried that input beforehand with probability at most \(q_H2^{-\secpar}\). 
The exposed \PINE view has the same distribution for both challenge clients because they use the same vector and \PINE parameters, and \PINE does not depend on the credential secret.

\textsf{Game 3}: \emph{(Remove the ciphertext plaintext).}
Replace the challenge ciphertext \((\gacc^\rho,\TraceHandle_b\TracePK^\rho)\) with an encryption of an independent random group element. The adversary is already given \(\TraceHandle_0\) and \(\TraceHandle_1\), so this is a chosen-plaintext threshold-ElGamal experiment in the presence of the sub-threshold tracing shares allowed by Definition~\ref{def:anonymity-game}. Any distinguisher between Games~2 and~3 therefore breaks the assumed threshold-ElGamal IND-CPA security.

\textsf{Game 4}: \emph{(Remove the round-pseudonym exponent).}
Replace \(\roundlink=\roundlinkbase^{\clientsecret_b}\) by a uniform element of \(\GOne\). Given \((\gacc,\gacc^{\clientsecret_i},\roundlinkbase)\), distinguishing \(\roundlinkbase^{\clientsecret_i}\) from random is a DDH problem. A standard two-step hybrid handles the two possible client secrets.

The directory handles and enrolment transcripts can also be simulated without knowing \(\clientsecret_i\). The reduction chooses \(\kappa\), sets
\[
  \IdentPK=\gacc^\kappa,
  \qquad
  \DirHandle_i=\TraceHandle_i^\kappa,
\]
and therefore computes each directory handle directly from the known tracing handle.

The enrolment transcript can also be simulated without knowing \(\clientsecret_i\). Fresh \(s_i\) makes \(C'_i\) uniform in \(\GOne\), so the reduction samples \(C'_i\) uniformly subject to \(\gone C'_i\neq1_{\GOne}\), chooses
\(e_i\in\Zp\setminus\{-x\}\), and sets
\(
  A_i=(\gone C'_i)^{1/(x+e_i)}.
\)
It then simulates the enrolment proof using Lemma~\ref{lem:enrol-pok}. Excluding the single invalid value changes the distribution by only \(1/p\). For every non-challenge context, the reduction chooses \(a_t\getsr\Zps\), sets
\[
  \roundlinkbase=\gacc^{a_t},
  \qquad
  \roundlink=\TraceHandle_i^{a_t},
\]
and therefore generates the pseudonym without knowing \(\clientsecret_i\). For the challenge context, it uses the DDH challenge base as the output of \(\HashG\). 
If the context is chosen adaptively, the reduction guesses which \(\HashG\) query will be used for the challenge, incurring only the standard polynomial loss. These values have the same distribution as in the real protocol. Moreover, the protocol publishes no \(\GTwo\) encoding of \(\clientsecret_i\) or \(\kappa\), so pairings do not provide an additional test linking the published handles.

In Game~4, the presentation, unseen digest, ciphertext, pseudonym, and exposed \PINE view are independent of \(b\), and \(\sid\) is a deterministic hash of the resulting statement.

Finally, \(\Pr[b'=b]=1/2\). Game~1 differs only by negligible random-oracle programming failure, Game~2 by at most \(q_H2^{-\secpar}\), and the remaining hops are indistinguishable under threshold-ElGamal IND-CPA and DDH. Therefore
\[
  \Adv^{\mathsf{anon}}_{\VeilFL,\mathcal A}
  =\negl(\secpar).
\]
\end{proof}

\subsection{Fault Non-Frameability} \label{app:frame}
This subsection defines the fault non-frameability experiment and proves Theorem~\ref{thm:nonframe} by analysing the possible forms of a confirmed fault.
\begin{definition}[Fault non-frameability experiment]\label{def:frame-game}
The experiment \(\mathsf{Exp}^{\mathsf{frame}}_{\VeilFL,\mathcal A}(\secpar)\) is defined as follows.

\begin{enumerate}
  \item The challenger runs setup and enrols one honest target client \(\TargetClient\) with secret \(\clientsecret^{\star}\) and tracing handle
  \[
    \TraceHandle^{\star}
    =\gacc^{\clientsecret^{\star}}.
  \]
  The adversary receives the public records, one aggregator's view, malicious-client state, and subthreshold committee shares. It does not receive the target credential or secret.

  \item Here, \(\mathsf{HonestSubmit}(\context,\qvec)\) runs Algorithm~\ref{alg:submit} for the honest target on \((\context,\qvec)\). The adversary may call \(\mathsf{HonestSubmit}(\context,\qvec)\) at most \(Q_h\) times. The oracle accepts only a vector within the announced encoded-integer norm bound and only if the target has not submitted in that context. It runs the honest client. If the \PINE prover aborts, the oracle returns \(\bot\) and records no submission. Otherwise, it delivers both private messages intact and records the resulting public part as authorised.

  \item The adversary controls all other clients and network scheduling and may submit arbitrary messages to the protocol. The two aggregators and the log follow their specified algorithms. Exact replays are ignored.

  \item The experiment returns \(1\) if the log confirms a rejection or duplicate record whose valid Tier~1 opening equals \(\TraceHandle^{\star}\). Otherwise, it returns \(0\).
\end{enumerate}

The adversary's advantage is 
\[
  \Adv^{\mathsf{frame}}_{\VeilFL,\mathcal A}(\secpar)
  :=
  \Pr\!\left[
    \mathsf{Exp}^{\mathsf{frame}}_{\VeilFL,\mathcal A}(\secpar)=1
  \right].
\]
\end{definition}

Every authorised target submission is complete, in bound, and unique in its context. A winning fault therefore uses an unauthorised public part, a duplicate record, or a rejection record for an authorised submission. Any confirmed record must also contain the required aggregator signatures; otherwise the adversary forges one.

\begin{proof}[Proof of Theorem~\ref{thm:nonframe}]
Assume that the experiment returns \(1\), and consider the first confirmed fault whose valid Tier~1 opening equals \(\TraceHandle^{\star}\).

\emph{(i) Unauthorised public part.}
Suppose the public part selected for opening was not returned by \(\mathsf{HonestSubmit}\). We reduce this event to discrete-log hardness in \(\GOne\). The reduction receives
\[
  \TraceHandle^{\star}
  =\gacc^{\clientsecret^{\star}}
\]
for uniformly random, unknown \(\clientsecret^{\star}\in\Zps\), and uses this element as the target tracing handle. The target credential and enrolment transcript are not revealed in this experiment.

The reduction answers each \(\HashG\) query with \(\gacc^a\) for fresh \(a\getsr\Zps\) and records \(a\). For an \(\mathsf{HonestSubmit}\) query in context \(t\), it uses the recorded \(a_t\) to set
\[
  \roundlinkbase_t=\gacc^{a_t},
  \qquad
  \roundlink_t=(\TraceHandle^{\star})^{a_t},
\]
and forms
\(
  (\CtOne,\CtTwo)
  =
  (\gacc^\rho,
   \TraceHandle^{\star}\TracePK^\rho).
\)
It generates the \PINE data honestly, samples \(\Abar,D\getsr\GOne\setminus\{1_{\GOne}\}\), sets \(\Bbar=\Abar^x\), and simulates \(\pi^{\mathsf{sub}}\) using Lemma~\ref{lem:submission-pok}. Thus, the target submissions have the honest distribution up to negligible random-oracle programming failure, without the reduction knowing \(\clientsecret^{\star}\).

A fault-producing unauthorised submission has a different submission identifier from every authorised target submission. Except on a collision in \(\Hash\), its submission-proof statement is therefore distinct from those used by the simulated honest submissions. The submission-proof extractor of Lemma~\ref{lem:submission-pok} can thus fork the Fiat-Shamir query for this statement while keeping the random-oracle points programmed for the simulated honest proofs fixed.

Extraction yields \(\widehat{\clientsecret}\) such that the tracing ciphertext encrypts \(\gacc^{\widehat{\clientsecret}}\). By Lemma~\ref{lem:share-pok} and correctness of threshold reconstruction, the Tier~1 opening equals this plaintext. Since the winning condition requires the opening to be \(\TraceHandle^{\star}\),
\[
  \gacc^{\widehat{\clientsecret}}
  =
  \TraceHandle^{\star}
  =
  \gacc^{\clientsecret^{\star}},
\]
and hence
\(
  \widehat{\clientsecret}
  =
  \clientsecret^{\star}\pmod p.
\)
The reduction therefore recovers \(\log_{\gacc}\TraceHandle^{\star}\), contradicting discrete-log hardness, which is implied by DDH hardness in \(\GOne\).

\emph{(ii) Duplicate record.}
The honest oracle creates at most one target submission per context. A duplicate therefore contains two distinct valid submissions with the same \((\context,\roundlink)\), at least one of which is unauthorised.
By Theorem~\ref{thm:binding}, the two submissions use the same client secret. Since the selected submission opens to \(\TraceHandle^{\star}=\gacc^{\clientsecret^{\star}}\), the unauthorised submission is tied to \(\clientsecret^{\star}\) and reduces to the preceding case. A replay does not create a duplicate because it has the same submission identifier.

\emph{(iii) Rejection of an authorised public part.}
Suppose the rejection record uses a public part returned by \(\mathsf{HonestSubmit}\). Such a submission is recorded only after successful \PINE proof generation for a vector within the announced bound, and both private messages are delivered intact. By Section~\ref{sec:pine-prelim}, the protocol-compliant verifiers then obtain the common accepting output, so no rejection record is formed.

If the network adversary drops or alters a private message, the submission is incomplete or fails its digest check and does not produce a rejection fault. Making a different message pass under the same advertised digest requires a collision in \(\Hash\).

Thus, a winning execution implies a failure of submission-proof knowledge soundness, discrete-log hardness, aggregator-signature unforgeability, collision resistance, or opening-share proof soundness, or a negligible random-oracle programming failure. Therefore
\[
  \Pr[\mathsf{frame}]
  \le \negl(\secpar).
\]
\end{proof}

\subsection{Controlled Opening}\label{app:opening-proof}
This subsection provides additional details for the correctness and privacy argument of Theorem~\ref{thm:opening}.
\paragraph{Proof Details}
The two interpolation equations in the main proof establish correctness of Tier~1 and Tier~2 reconstruction. Lemma~\ref{lem:share-pok} ensures that a malformed opening share cannot pass under an honest member's registered verification value, except with negligible probability.

For privacy before Tier~1, consider a tracing handle that has not already been revealed by an authorised opening. First apply the presentation-simulation and round-pseudonym DDH hybrids from Theorem~\ref{thm:anonymity}. After these hybrids, the tracing ciphertext is the remaining target-secret-dependent value, and threshold-ElGamal IND-CPA security hides its plaintext from a sub-threshold tracing coalition. Sampling \(\rho\) from \(\Zps\) rather than \(\Zp\) changes the standard ElGamal distribution by at most \(1/p\).

After Tier~1, the adversary knows
\(
  (\gacc,\IdentPK=\gacc^\kappa,
   \TraceHandle=\gacc^{\clientsecret}).
\)
The target's auxiliary records remain simulatable. For each fresh round the reduction programs \(\roundlinkbase=\gacc^{a_t}\), computes \(\roundlink=\TraceHandle^{a_t}\), simulates the randomised BBS presentation, and encrypts the known group element \(\TraceHandle\). Computing the previously unknown value
\[
  \DirHandle=\TraceHandle^\kappa
\]
is therefore CDH, while deciding whether a known candidate is the corresponding directory handle is DDH. Subthreshold identification shares do not change these conclusions under the assumed threshold-key security.

Non-target openings do not require the target CDH value. The reduction generates the tracing key and can therefore answer non-target Tier~1 openings. 
For Tier~2, if a non-target submission has known or extracted secret \(\clientsecret_i\), it computes
\[
  Z_{i,j}=\IdentPK_j^{\clientsecret_i}
\]
and simulates the corresponding Chaum-Pedersen proof. Honest non-target secrets are chosen by the reduction, while knowledge soundness of Lemma~\ref{lem:submission-pok} supplies the witness for an accepted malicious non-target submission. Thus, only the unopened target directory handle carries the CDH challenge.

Finally, the requirement for \(\faultthreshold\) linked faults is enforced by honest identification members before releasing their shares. A coalition below the identification threshold cannot assemble an opening before this condition is met; a fully corrupted quorum is outside the threat model.


\cleardoublepage
\end{document}